\documentclass[preprint2]{aastex}
\usepackage{longtable}

\slugcomment{Not to appear in Nonlearned J., 45.}

\shorttitle{XMM-Newton Observations of Luminous Sources in the Nearby Galaxies}
\shortauthors{ Akyuz et al.}

\begin{document}


\title{{\it XMM-Newton} Observations of Luminous Sources in the Nearby Galaxies: NGC 4395, NGC 4736, and NGC 4258}


\author{A. Akyuz\altaffilmark{1}, S. Kayaci\altaffilmark{2}, H. Avdan\altaffilmark{1}, M. E. Ozel\altaffilmark{3}, E. Sonbas\altaffilmark{4}, S. Balman\altaffilmark{5}} 

\affil{$^1$University of Cukurova, Department of Physics, 01330 Adana, Turkey}
\affil{$^2$University of Erciyes, Department of Astronomy, Kayseri, Turkey}
\affil{$^3$\c{C}a\u{g} University, Faculty of Arts and Sciences, 33800 Yenice, Tarsus, Mersin, Turkey}
\affil{$^4$University of Ad$\i$yaman, Department of Physics, 02040 Ad$\i$yaman, Turkey }
\affil{$^5$Middle East Technical University, Dept. of Physics, 06531 Ankara,  Turkey}
\email{aakyuz@cu.edu.tr}





\begin{abstract}
We present results from a study of the non-nuclear discrete sources in a sample of three  nearby  spiral galaxies (NGC 4395, NGC 4736, and NGC 4258) based on {\it XMM-Newton} archival data supplemented with {\it Chandra} data for spectral and timing analyses. A total of 75 X-ray sources has been detected within the {\it D$_{25}$} regions of the target galaxies. The large collecting area of {\it XMM-Newton} makes the statistics sufficient to obtain spectral fitting for 16 (about 20$\%$) of these sources. Compiling the extensive archival exposures available, we were able to obtain the detailed spectral shapes of diverse classes of point sources. We have also studied temporal properties of these luminous sources. 11 of them are found to show short-term (less than 80 ks) variation while 8 of them show long-term variation within factors of $\sim$ 2 to 5 during a time interval of $\sim$ 2 to 12 years. Timing analysis provides strong evidence that most of these sources are accreting X-ray binary (XRB) systems. One source that has properties different than others was suspected to be a Supernova Remnant (SNR), and our follow-up optical observation confirmed it. Our results indicate that sources within the three nearby galaxies are showing a variety of source populations, including several Ultraluminous X-Ray Sources (ULXs), X-ray binaries (XRBs), transients together with a Super Soft Source (SSS) and a background Active Galactic Nucleus (AGN) candidate.
\end{abstract}


\section{Introduction}

The increased imaging and energy resolution capabilities of the new generation X-ray observatories like {\it XMM-Newton} and {\it Chandra}, provide a better understanding of the X-ray emission from  nearby galaxies. A combination of a number of discrete sources, hot interstellar gas, and a probable active galactic nucleus are thought to produce the observed X-ray emission from such galaxies (Fabbiano, 1989). Large number of discrete sources have already been detected by {\it XMM-Newton} and {\it Chandra} in the disks of nearby galaxies such as M 31 (Pietsch 2005; Kong et al. 2002a, 2003b), M 33 (Pietsch 2004; Plucinsky et al. 2008) M 81 (Swartz et al. 2003), and M 101 (Pence et al. 2001; Jenkins et al. 2004). These sources are typically high and low mass X-ray binaries (XRBs), supernova remnants (SNRs), bright super soft sources (SSS), and a number of very luminous sources, so-called ultra-luminous X-ray sources (ULXs).

For a study of discrete X-ray source population in the nearby spiral galaxies, a sample of three of them (NGC 4395, NGC 4736, and NGC 4258) were chosen for this work. All three have a low activity Seyfert nucleus. Their low foreground absorption (1.20-1.43)$\times$10$^{20}$ cm$^{-2}$ (Dickey $\&$ Lockman 1990) also made them good targets for exploring such discrete X-ray sources. 
 
NGC 4395 is classified as one of the nearest and least luminous (L$_{Bol}$ $\sim$ 5$\times$10$^{40}$ erg s$^{-1}$)  Type 1 Seyfert galaxies (Filippenko $\&$ Sargent 1989). This dwarf galaxy appears to harbor a  central black-hole, likely to be in the $\sim$ 10$^{4}$ -10$^{5}$ M$_{\odot}$ range, significantly lower than more luminous Seyferts with black hole masses typically in the range $\sim$ 10$^{6}$ -10$^{8}$ M$_{\odot}$ (Vaughan et al. 2005). Its nucleus has been the subject of several X-ray studies since the {\it ROSAT} observations (Moran et al. 1999; Lira et. al 1999). They have reported that its nuclear X-ray source displays a large amplitude variability on short time scales and has a soft X-ray spectrum. Also, five X-ray sources, one being significantly brighter than nucleus,  were detected within $\sim$3$\arcmin$ of its nucleus. Later, observations by {\it Chandra} and {\it XMM-Newton} confirmed the variable nature of its active nucleus (Moran et al. 2005; Vaughan et al. 2005). 

Our next target, NGC 4736 (M94), is classified as a low luminosity  galaxy with the closest sample of a low ionization nuclear emission-line (LINER) nucleus (Flippenko $\&$ Sargent 1985). Two  bright sources were detected by {\it ROSAT} observations in its nuclear area (Cui et al. 1997). One of them was consistent with the nucleus of the galaxy and  almost one-third of the total observed flux in (0.1-2) keV arose as the emission from the compact source in the center with an extended distribution of hot gas. 12 discrete sources (including galactic nucleus itself) were identified by {\it ROSAT HRI} data in its optical disk and ring regions (Roberts et al. 1999). The galaxy was also observed with the {\it BeppoSAX } in (0.1 - 100 keV) band and the {\it Chandra ACIS-S} for  a high resolution study of its nuclear region (Pellegrini et al. 2002).  They concluded that the LINER activity was due to a low luminosity AGN at the center. Eracleous et al. (2002) had already examined the high concentration of luminous sources in the nucleus of NGC 4736 by {\it Chandra}. They suggested that the galaxy was in a recent starburst phase where the emission was dominated by dense clusters of X-ray binaries.

Our third target, NGC 4258 (M106) is a bright nearby Type 1.9 Seyfert galaxy. It is also known by its anomalous arms discovered by H$\alpha$ imaging (Courtes et al. 1993). Water maser line emission of rotating gas near its center indicated a mass of 3.6$\times$10$^{7}$M$_{\odot}$ in a  disk of radius ranging from 0.12 to 0.25 pc (Miyoshi et al. 1995). {\it ROSAT} observations of NGC 4258  resolved  the anomalous spiral arms forming the boundary of diffuse emission. 15 X-ray point sources associated with the galaxy were identified by {\it ROSAT} observations (Pietsch et al. 1994; Cecil et al. 1995; Vogler $\&$ Pietsch 1999). They could not detect the Seyfert nucleus of the galaxy as a point source. The first detection of active nucleus was reported by  {\it ASCA} observations. It was modeled  by an absorbed power-law spectrum of photon index $\Gamma$ $\sim$1.8. The unabsorbed  (2-10) keV   luminosity of the source was {\it L$_{X}$}= 4$\times$10$^{40}$ ergs$^{-1}$ obscured by a column density of N$_{H}$ $\sim$ 1.5$\times$10$^{23}$ cm$^{-2}$. An Fe K$\alpha$ emission line was detected with an equivalent width 0.25$\pm$0.10 keV (Makishima et al. 1994). 

NGC 4258 was also observed by {\it XMM-Newton}, which showed a hard nuclear point source that could be modeled by a highly absorbed power-law spectrum with no narrow Fe K$\alpha$ emission (Pietsch $\&$ Read 2002). A series of observations were also made by {\it Chandra} to study its various properties (Young $\&$ Wilson 2004). They examined the spectrum of the low-luminosity active galactic nucleus, which was well described by a hard X-ray power-law and a variable luminosity with a constant thermal soft X-ray component. They also discussed the first detection of extragalactic Fe absorption line at 6.9 keV,  with a very inclined accretion disk varying on a timescale of 6 ks. Yang et al. (2007) studied {\it Chandra} and  {\it XMM - Newton} spectra of its anomalous arms and active jets. They confirmed that the X-ray emission from these arms has a thermal nature, suggesting that currently active jets could be responsible for heating the gas in the arms. 

In the present work, we will investigate the spectral and timing properties of the bright point-like non-nuclear sources in these three galaxies, first as observed by  {\it XMM-Newton} with a more extended database compared with the previous studies and later, by  {\it Chandra} to supplement their spectral and timing properties.
  General properties of the  galaxies are listed in Table 1. We will restrict ourselves to the off-nucleus luminous sources with sufficient signal to noise ratio (S/N) to perform  spectral fitting. Source locations with respect to the galaxy centers and their cross identifications (if any) have been listed in Table 2.

 The organization of paper will be as follows; In Section 2,  we outline the data reduction and source selection criteria. In Section 3, the spectral fitting and timing analysis techniques are described and, in Section 4, detailed discussion of spectral and temporal results for individual sources in the galaxies are presented. The overall properties of their  X-ray emission are discussed in Section 5.       


\section{Observations and Data reduction}

\subsection{ The {\it XMM-Newton} observation}
The data used for the analysis were retrieved from the {\it XMM-Newton} public archive. {\it XMM-Newton} Observatory (Jansen et al. 2001) has three EPIC cameras at its focal plane:  one uses the pn-CCD (Str\"{u}der et al. 2001) and the other two MOS-CCD detectors (Turner et al. 2001). Instruments were operated in the full-frame imaging mode and used thin or medium filters for pn and MOS. We analyzed the pipeline-processed data by using the standard software tools of {\it XMM-Newton Science Analysis System (XMM-SAS) v 10.0} (Gabriel et al. 2004). In the {\it SAS},  X-ray events corresponding to PATTERN$\leq$ 4 (singles and doubles) were then selected with FLAG==0 option for the pn camera, PATTERN$\leq$ 12 were used for the MOS cameras. Source detection was performed using the standard maximum likelihood technique as implemented by  {\it SAS} tool {\it edetect$_{-}$chain}.  Each source inside the {\it D$_{25}$} ellipse was checked to exclude false or doubtful sources, after the automatic selection procedure in {\it SAS}. 
The log of observations, including the  dates, IDs, exposure times, and the type of EPIC camera for each target, are given in Table 3. 

In the analysis, we simultaneously used data from the EPIC pn and the EPIC MOS for each galaxy to determine the  spectra  of the sources  (with the exception of the MOS data for  NGC 4395, which has poor statistics). Note also that the EPIC pn  has more source counts due to its better sensitivity when compared with the MOS detectors. Figures 1a, 2a, and 3a  show the combined exposure-corrected EPIC pn, MOS1 and MOS2 true color images of the presently analyzed archival data for the sample galaxies. 
Inclination corrected {\it D$_{25}$} ellipses are also shown for comparison and identification. The detected point sources inside a {\it D$_{25}$} ellipse were specified as off-nuclear objects. We numbered them as XMM-n, n representing source number with decreasing pn count rate. Figures 1b, 2b, and 3b show  {\it XMM-Newton} point source positions overlaid on a DSS1 optical blue image of each galaxy. Detected X-ray source parameters in each galaxy are summarized in Tables 4, 5, and 6 respectively. In each table the source ID (col.1), its position (col.2 \& 3), the likelihood of existence (where a likelihood value of 10 corresponding to a detection probability above 4$\sigma$) (col.4), integrated (pn + MOS1 + MOS2) count rates with errors (col.5), flux with the error (col.6), and luminosity with the error (col.7) are given in the (0.2-12) keV band.  The luminosities were calculated
from the flux values using the distances indicated in Table 1. 

\subsection{The {\it Chandra} observation }

To study the spectral and temporal properties of the non-nuclear point sources, we complemented our {\it XMM-Newton} analysis with the {\it Chandra} data of each target galaxy. A separate  observation log of the {\it Chandra} data is given in Table 7 for our galaxies. We retrieved the data from the public archive and analyzed with {\it CIAO v4.3 } software package using the CALDB v4.4.2 calibration database. All data were filtered to 0.3 - 10 keV energy range. The detection of the X-ray point sources was performed using the WAVDETECT task within the same package. We have run the task over the 1, 2, 4, 8 and 16-pixel wavelet scales, since these scales are appropriate for detecting point-like sources. After a visual examination for possible false detections, we have only taken into account the sources with detection significance of 3.5$\sigma$ or better and with net counts higher than 6. 

 It is well known that, {\it Chandra} has significantly better angular resolution than {\it XMM-Newton}. As a result, the detected  {\it XMM-Newton}  point sources (one or two) in the central region of galaxies may have overlapped several {\it Chandra} point sources. Since in the present analyses,  we examined only the non-nuclear sources, the sources located in the central regions of 0.3$\arcmin$ radius were excluded  in all three galaxies. 
 
  We have started  our {\it Chandra} data analysis with NGC 4395 and the detected 11 point sources in its {\it D$_{25}$} region. Only one source in the  central region of the galaxy has been excluded. From the remaining 10,  several have coincided  with corresponding {\it XMM-Newton} sources. Four of them  were examined individually in Section 5.1 due to sufficient statistics provide by {\it XMM-Newton} observations.   These four {\it Chandra} source positions are offset within 2$\arcsec$ of {\it XMM-Newton} source positions center in the galaxy.

When the same procedure was carried out for the other two galaxies we obtained the following:  For NGC 4736, 43 point sources in the {\it D$_{25}$} region were  detected.
15 sources which were in the central region  have been excluded. From the 
remaining 28 several overlapped by {\it XMM-Newton} sources. Among are them, three sources  coincided with  {\it XMM-Newton}
sources, which were examined individually in Section 5.2. {\it XMM-Newton} and {\it Chandra} source positions of these three  agree within 2$\arcsec$ in NGC 4736.

For NGC 4258, 42 point sources  were  detected in the {\it D$_{25}$} region. Five point sources in the central region have been removed. Among the remaining,  six sources have coincided  with corresponding {\it XMM-Newton} sources, and they were examined individually in Section 5.3. 
 {\it XMM-Newton} and {\it Chandra} source positions of these six  did agree  within  $\sim$ $< $1$\arcsec$ (except XMM-8 with two {\it Chandra} sources within its 15$\arcsec$ {\it XMM-Newton} error circle).

\section{Spectral Analysis}

To perform a spectral analysis for each luminous source, corresponding data  were extracted from a circular region with a radius (15$\arcsec$ - 20$\arcsec$) for pn, MOS1 and MOS2 depending on the presence of nearby sources. Background events were extracted from a source-free zone normalized to the  source extraction area. 

The source spectra  with the instrument responses and ancillary files were generated using the {\it SAS v 10.0}. The  spectral analyses of the X-ray data have been performed using {\it XSPEC v.12.5}.  We grouped the pn and MOS spectral energy channels  to have at least 20-30 counts per bin for a good statistical quality in the spectra. Initially we fitted the spectra with  four different  one-component models: blackbody (BBODY), power-law (PL), diskblackbody (DISKBB), and bremsstrahlung (BREMSS). In addition, we applied the following two-component models : powerlaw+blackbody (PL+BBODY), powerlaw+disk blackbody (PL+DISKBB), and powerlaw+mekal (PL+MEKAL). We also used a photoelectric absorption model as part of the fit. The fits were applied to data in the 0.2 - 10 keV range.

The EPIC pn and MOS spectra  were fitted simultaneously for each object only in galaxies NGC 4736 and NGC 4258. For NGC 4395 we used only the EPIC pn data due to low statistical quality of the MOS data. We used a constant parameter which was set free to account for the normalization differences in the flux calibration of the three EPIC cameras. 

To compare the spectral shapes of sources between {\it XMM-Newton} and {\it Chandra} observations, we fitted the same spectral models to all data. Each data set was chosen based on a longer exposure time to provide sufficient counts for spectral fitting.  Standard {\it CIAO} routines were used  to reprocess event lists; then source and background spectra were extracted to construct so called the RMF and ARF files. 
We have fitted the data of 13 {\it Chandra} sources in the three galaxies, however, only two of them (XMM-3 and XMM-6 in NGC 4258) were above the threshold ($>$ 300 counts) for spectral analysis. Our  results are consistent with the {\it XMM-Newton} analysis therefore, we have discussed only the spectral parameters of these two sources in NGC 4258 in Section 5.3. 

\section{Timing Analysis}

X-ray variability of sources in all three  galaxies were tested by deriving their short-term (based on the observation with the longest exposure available) and long-term  light curves. For short-term variability, the data from EPIC pn and MOS cameras were added to improve S/N. After excluding the high background flaring,  the corrected exposure times were binned in the range from 200 to 4500 s and each bin had at least 20 counts after background subtraction.

We applied a $\chi^{2}$ test to search for large amplitude variations with respect to the constant count rate hypothesis. Short-term variability of point sources for each galaxy in our list can be followed from Figures 4, 5, and 6 respectively. In Table 8 we present the results of $\chi^{2}$ tests together with {\it P(var)} the probability values for those cases exceeding the 95\% taken as the  limit for variability. Results are such that, in NGC 4395 one out of 5, in NGC 4736 one out of 3 of the sources are variable while in NGC 4258 variable sources increased to 7 out of  8. Kolmogorov-Smirnov (K-S) tests were also applied to search for small amplitude variations in the source count rates; however, nothing significant was found.

For long-term variability, in addition to   {\it XMM-Newton} and {\it Chandra}, {\it ROSAT} data from the literature were also added to the analysis, all in the  0.5-2 keV energy range. In this application, To calculate {\it XMM-Newton} fluxes, data were fitted with their best fitting models and  fluxes calculated after the spectra of the sources determined.

For {\it Chandra} source fluxes, even though the data for 11 out of 13 sources were below the threshold (as noted above), for comparison purposes, we have tried to fit the data with the same best fitting   {\it XMM-Newton} model with no constrained on {\it Chandra} data fit parameters. 
  We have succeeded in getting flux values for only two sources in NGC 4395 (XMM-5 and XMM-6), two sources in NGC 4736 (XMM-2 and XMM-12), and five sources in NGC 4258 (XMM-3, XMM-6, XMM-16, XMM-17, XMM-21) within the same energy range (0.5-2) keV. In our timing analysis, we used the longer exposure time if the data were obtained within a year. 

 To derive fluxes for  {\it ROSAT} observations, count rates from available literature were used. The absorbed PL continuum slope value was taken from the {\it XMM-Newton} observation and the standard {\it WebPIMMS} was used to calculate flux. 
 
 Resulting long-term light curves for sources in all three galaxies are given in Figures 7a, 7b, and 7c.  For the transient ULX source in NGC 4258, short and long-term light curves are shown in Figures 8a, 8b. Here, the total number of points of observation for the long time interval (18 month to 12 years) looks rather sparse. However, there are noticeable indications of long term variability for a number of sources studied. Our variability search for individual sources in each galaxy will be futher discussed in the following sections.
 
\section {Individual Source Properties}
In this section, we will discuss observations of each galaxy individually, highlighting the spectral and timing properties of X-ray luminous point sources in turn.

{\bf\subsection{NGC 4395}}
This galaxy was observed in 2002 and 2003 (see Table 2). Within the {\it D$_{25}$} region, we have detected 29 point sources as given in Table 3. Our detection limiting flux in the 0.2-12 keV  energy band is $\sim$ 1.2 $\times$10$^{-15}$ erg cm$^{-2}$s$^{-1}$ for sources inside the optical disk.
The best fitting model spectra for 5 sources were obtained by using only the pn data since the MOS data had rather poor statistics. In our further analysis, we obtained the best-fitting spectral parameters for one-component and two-component models for the 5 sources. Our results are summarized in Tables 9 and 10 respectively, together with unabsorbed fluxes and luminosities in the energy range 0.2-10 keV. The best fitting model spectra of these sources are given in Figure 9. 
We will elaborate on their individual spectral characteristics below: \\

{\bf XMM-2}: This is the brightest object in the galaxy and is about  2.9$\arcmin$  away from the galaxy center. It is located nearby a giant H II region complex. The distance between the position of the source XMM-2 and the center of the near HII region  is no less than $\sim$ 500 pc. This calculation takes into account a deprojected distance using the inclination of the galaxy. (Distribution of H II regions in this galaxy was studied by  Cedres \& Cepa (2002) in detail).  The source spectrum obtained using  {\it XMM-Newton} data is not adequately fitted with any one-component model. The best-fitting two-component model is an absorbed  PL + DISKBB ($\Gamma$ = 3.11, {\it T$_{in}$} = 0.19 keV).  But a PL+MEKAL model gives almost identical fit ($\chi^{2}_{\nu}$= 1.2) with similar parameters.   It has an unabsorbed luminosity of {\it L$_{X}$} = 2.2 $\times$ th10$^{39}$ erg s${-1}$ which is within the standard ULX limit.   Both the short-term and long-term light curves do not  show evidence of variability. However, its spectral shape and high luminosity lead us to conclude that it is a ULX. This source was claimed earlier to be a ULX by Stobbart et al. (2006).  XMM-2 was also defined as a ULX with 
the best-fitted two-component model, BB+PL ($\Gamma$ = 3.44, \textit{kT} = 0.14 keV), using 2000 and 2002 archival data by Winter, Mushotzky $\&$ Reynolds (2006) (hereafter WMR2006). We think that this ULX source needs deeper observations to decide on  its spectral and temporal characteristics better. \\

{\bf XMM-5}: This is the most notable source in this galaxy. We derived a source spectrum and fitted this spectrum with a VPSHOCK model (another {\it XSPEC} model) of a non-equilibrium, ionization plasma emission as suitable for SNRs (Borkowski et al. 1996). The resulting spectral parameters are \textit{kT} = 3.98 keV with an unabsorbed luminosity of {\it L$_{X}$}  $\sim$ 1.1 $\times$10$^{38}$ erg s$^{-1}$.  The light curve for XMM-5 shows no variability on both long and short time-scales; this is  expected since it is an SNR candidate, the only one in our entire source catalog with a possible extended nature.

Upon this unique possibility, we also conducted an optical spectroscopic observation to confirm and to clarify its nature. In order to differentiate SNRs from typical H II regions, we used the well known criterion [S II]/H$\alpha$ $\geq$ 0.4 proposed by Mathewson $\&$ Clark (1973). The spectral data were obtained by TFOSC (TUG Faint Object Spectrograph and Camera) instrument mounted on RTT150 (Russian-Turkish Telescope at Antalya, Turkey) in April 2009. Optical spectrum of the source is given in Fig. 10.  In Table 11 observed line intensities, the E$_{(B-V)}$, and H$\alpha$ intensity values are listed. As these optical observations reveal, this source is indeed an SNR rather than a ULX claimed by Swartz et al. (2004). In a more recent study of this galaxy, SNR nature of XMM-5 was also discussed by Leonidaki et al. (2010) using a higher resolution {\it Chandra} archival data. \\

{\bf XMM-6 and XMM-10}: The spectra of these sources are best fitted with an absorbed PL model with a photon index of  $\Gamma$ = 1.42 and $\Gamma$ = 1.86 respectively. Quite similar luminosity values of {\it L$_{X}$} = 2.6 $\times$10$^{38}$ erg s$^{-1}$ and $\sim$ 1 $\times$10$^{38}$ erg s$^{-1}$ are also observed respectively. These two sources are at different distances from the central bulge of the galaxy, however we think that XMM-6 has a noticeable contribution from the galactic bulge material, which may explain the differences in spectral shapes and luminosities. We detected no definite variation in long-term light curve of XMM-6, but  its short-term light curve satisfies our criteria of variability. XMM-10,  does not show any definite indication of short-term variability although its long-term light curve does show some variability (a factor of $\sim$2). We can conclude that both sources are XRBs.     \\

{\bf XMM-23}: This source is located about 2$^{\prime}$  south of the central bulge at a deprojected distance of about 2.4 kpc. It displays a very soft spectrum that fits well with an absorbed cool BBODY (\textit{kT}  $\sim$ 60 eV,  $\chi^{2}_{\nu}$ $\sim$ 1.3) model with an unabsorbed luminosity of {\it L$_{X}$} $\sim$ 6$\times$10$^{37}$ erg s$^{-1}$. Its short-term light curve does not show definite variability, but  long-term light curve shows variations  spanning $\sim$ 2 year.  With this cool temperature, similar to those found in other galaxies (Swartz et al. 2002, Di Stefano et al. 2004), it may be classified as an SSS. They are known as  cataclysmic binaries accreting at high rate where hydrogen on the surface of the white dwarf is burned leading to an Eddington luminosity (Di Stefano $\&$ Kong, 2003). 

{\bf\subsection{NGC 4736}}
We determined 21 point sources within its  {\it D$_{25}$} boundaries. These sources and relevant parameters are given in Table 4. 
The detection limiting flux in the 0.2-12 keV  energy band is $\sim$ 2 $\times$10$^{-15}$ erg cm$^{-2}$s$^{-1}$ for sources inside the optical disk.
We were able to obtain X-ray spectra of 4 of them. These sources have not been studied in the X-rays before. The spectral parameters for these sources including unabsorbed flux and luminosities are given in Tables 12-13. The fitted spectra for these sources are plotted in Figure 11.  For the individual sources we note the following: \\

{\bf XMM-2}: This source is located in the disk of the galaxy. A one-component BREMSS model fit to data yields a good fit with a $\chi^{2}_{\nu}$$\sim$ 1; however, the fitted temperature (\textit{kT} = $\sim$ 3.1 keV) is rather high. A two-component model PL+DISKBB on the other hand gives a good fit with the same $\chi^{2}_{\nu}$ and lower value of \textit{kT} $\sim$ 0.75 keV.  In the earlier work by WMR2006 using archival data of {\it XMM-Newton}, this source was not detected; however, we were able to find the source in the next data set about one month later (see Table 3). Its short-term light curve (in Fig. 5) shows episodes of  significant variability.  In the long-term, it has brightened a factor of $\sim$ 25  in the {\it XMM-Newton} observations during a $>$ 6 year period.  It has a high unabsorbed luminosity  of \textit{L$_{X}$} = 1.95$\times$10$^{39}$ erg s$^{-1}$ revealing that  this source can be a transient ULX. \\

{\bf XMM-12}: This source is located away from the disk of the galaxy and was  detected earlier by  {\it ROSAT} (Roberts et al. 1999). In our analysis its spectrum  is adequately fitted by the one-component DISKBB model ({\it T$_{in}$} = 1.44 keV) with an unabsorbed luminosity of {\it L$_{X}$} = 0.7$\times$10$^{38}$ erg s$^{-1}$.  Although it does not show any short-term variability, in the long-term  a variability  within a factor of 3 can not be excluded.  It could be an XRB source.\\

{\bf XMM-18}: This source was first observed by {\it ROSAT} (Roberts et al. 1999). It is  located  $\sim$2.1$^{\prime}$  away from the central region of galaxy. The source spectrum is best fitted by a PL ($\Gamma$ = 1.39) with an unabsorbed luminosity of L$_{X}$ = 0.7 $\times$10$^{38}$ erg s$^{-1}$.  While no significant variability is found in the  short-term, in the long-term a variability pattern of up to a factor of 3 is noticeable.  We suggest that this source is probably  an XRB. \\

{\bf\subsection{NGC 4258}}

We detected 24 sources inside the {\it D$_{25}$} region of this galaxy (see Table 5). Our detection limiting flux is $\sim$ 5 $\times$10$^{-15}$ erg cm$^{-2}$s$^{-1}$ in the 0.2-12 keV  energy band for sources inside the optical disk. More than half of these sources have high luminosity ({\it L$_{X}$} $>$ 10$^{38}$ ergs$^{-1}$). We derived one and two-component best-fitting spectral model parameters for the sources as given in Tables 14-15. The spectra the  best fitting models for 8 of these sources are given in Figure 12. Individual source characteristics are given in the following: \\

{\bf XMM-2}: We find that this source  is a transient ULX. Earlier {\it XMM-Newton} data was already interpreted by WMR2006 as a transient. The source has been detected for the first time in 2002  (ObsID 0059140901) and was not detected in other available observations. The sharp increase in count rates in 2002 data compared with previous (2000, 2001) and later (2006) upper limit leads us to firmly conclude its transient nature.  The source also shows short-term variability. Its spectrum is best described by DISKBB ({\it T$_{in}$}= 1.43 keV) with a $\chi^{2}_{\nu}$ value of 1.33 and  a high unabsorbed luminosity, {\it L$_{X}$} = 1.5 $\times$10$^{39}$ erg s$^{-1}$. The spectral fit is not improved  with an additional component. Our luminosity estimation for a one-component model is consistent with a ULX. \\

{\bf XMM-3}: The location of this source is quite far from the disk (2.1$^{\prime}$). This source was first detected by {\it ROSAT} (Pietsch et al 1994). It was also  identified as a ULX by {\it XMM-Newton} (WMR2006)  and  {\it Chandra} (Swartz et al. 2004). According to WMR2006, the source is best fitted with a  PL+ BB model (\textit{kT} = 0.78 keV and  $\Gamma$ = 2.02) with a $\chi^{2}_{\nu}$ value of 1.2 for the data in 2002. This  source was identified  by {\it Chandra} as (CXOU J121857.8+471607) with an absorbed PL model ($\Gamma$ = 2). 


In our {\it XMM-Newton} analysis, the source spectrum is best fitted with a one component absorbed DISKBB (T$_{in}$= 1.28 keV) with a rather high unabsorbed luminosity of {\it L$_{X}$} = 1.9 $\times$10$^{39}$ erg s$^{-1}$. Our analysis of the {\it Chandra} data for this source also show that its spectrum is best fitted with a hot DISKBB model, (T$_{in}$= 1.31 keV) with a higher unabsorbed luminosity of {\it L$_{X}$} = 2.8 $\times$10$^{39}$ erg s$^{-1}$.
Its short term light curve shows a significant variation, also its long-term light curve  shows variation with a factor of   $\sim$ 4. The high source luminosity, leads us to conclude that it is a ULX. We also note that the present analysis  is based on a longer observation time with much better statistics than  all earlier results.   \\

{\bf XMM-6}: This source is located 2.4$^{\prime}$ away from the galactic center. Its spectrum is well fitted with an absorbed PL ($\Gamma$ = 1.87) model. It has an unabsorbed luminosity of {\it L$_{X}$} = 1.5$\times$10$^{39}$ erg s$^{-1}$. A single component model fitted to the {\it Chandra} data gives almost the same parameter ($\Gamma$ = 1.87), and a similarly high luminosity  $\sim$ 1$\times$10$^{39}$ erg s$^{-1}$  helping to identify the source as an ULX (Shwartz et al. 2004). Our {\it Chandra} data analysis for this source also gives the same best fitted model (PL) and parameter  with almost the same  high luminosity value.

Long term light curve of XMM-6 shows  a factor of $\sim$ 3 variability between the intervals, and there is also strong evidence for short-term variability. This raises the possibility that it could be an accreting XRB in the galaxy. Interestingly, however, the same source also coincides with a blue stellar counterpart from DSS2 image making it a likely background AGN candidate. True nature of this source  requires further study. \\ 

{\bf XMM-8}: This source is best described by a two-component model. The spectral fit is improved over that of a simple PL ($\Gamma$ = 2.1) fit by  addition of a MEKAL component ($\Gamma$ = 1.95 and \textit{kT} = 0.57 keV).  It has a high unabsorbed luminosity of  {\it L$_{X}$} = 1.2 $\times$10$^{39}$ erg s$^{-1}$. To reveal its nature  we also analyzed the {\it Chandra} archival data. Two different sources can be resolved within the same position by the better angular resolution of {\it Chandra}. However, due to the limited amount of data, it was not possible for us to carry out a satisfactory spectral analysis for these two separate sources. In the same neighborhood, two radio SNR candidates were identified by Hyman et al. (2001) (sources 10 and 12, Table 1)  but none coinciding with our Chandra sources (angular separations $\sim$ 6$\arcsec$ and $\sim$ 16$\arcsec$ respectively). Both long-term and short-term light curves of data from the source region  do  show  evidences of sometimes erratic variability. The spectral shape discussed above leads us to tentatively conclude that, the source region could harbor an XRB. However, further long duration observations with better angular resolution are surely needed to clarify the nature of the source(s). \\  

{\bf XMM-10}: This source is located far away  (6.4$^{\prime}$) from the central body of the galaxy. Its spectrum is best fitted with an absorbed PL model ($\Gamma$ = 2.8, $\chi^{2}_{\nu}$= 0.9). The unabsorbed luminosity in this observation is {\it L$_{X}$} = 9.2$\times$10$^{38}$ erg s$^{-1}$. Its short-term light curve shows no strong indication of variability. However, its long-term light curve do show a variability (an uninterrupted increase) by  a factor of at least 5 over the time interval of a decade. This source could be an XRB. \\

{\bf XMM-16}: This source gives a  best  spectral fit with a PL model ($\Gamma$ = 2.20). The unabsorbed X-ray luminosity is {\it L$_{X}$} = 7.2$\times$10$^{38}$ erg s$^{-1}$.  Short-term light curve shows significant variability and the long-term data  shows evidence of variability within a factor   of $\sim$ 4 after 2002 in the next $\sim$ 4 years. The spectral parameters and variation of  the source  suggest  that this may also be a candidate XRB. \\

{\bf XMM-17}:  This source is located $\sim$ 2$^{\prime}$ away from main body of the galaxy and its spectrum is described by a PL with $\Gamma$ $\sim$ 2 with an unabsorbed luminosity of {\it L$_{X}$} = 3.3$\times$10$^{38}$ erg s$^{-1}$.  Its short-term light curve shows significant  level of variability, while its long-term data show  a factor of $\sim$ 3 variation between observations. This source may also be a candidate XRB. \\  

{\bf XMM-21}: In our analysis, this source was best fitted with a one-component absorbed PL model ($\Gamma$ $\sim$ 2). Both its short-term  and long-term data show  strong evidence for variability. The source has an  unabsorbed luminosity of {\it  L$_{X}$} = 2.0$\times$10$^{38}$ erg s$^{-1}$ which suggest that this source is likely to be an XRB. However in the list of WMR2006  it was  defined as a ULX. Its spectrum was fitted with a PL and a BREMSS equally well at a luminosity {\it L$_{X}$} = 1.2$\times$10$^{39}$ erg s$^{-1}$. Therefore, further observations are needed to clarify its true nature.\\

\section{Discussion and Conclusions}

In the above sections, we presented spectral and temporal  analyses on the non-nuclear  X-ray point sources in the three nearby galaxies, NGC 4395, NGC 4736, and NGC 4258 selected on the basis of their low hydrogen column densities  with low activity Seyfert nuclei. Our main conclusions can be noted as follows: 

(1) Total number of point sources in the {\it D$_{25}$}  area of each of these galaxies are similar (29 for NGC 4395, 21 for NGC 4736, and 23 for NGC4258). Data at hand however, allows spectral analysis for only a smaller number (a total of 16 out of 74) of these sources. Luminosity for these sources fall into the range (0.4 - 22.7)$\times$10$^{38}$ erg s$^{-1}$. 14 of these 16 sources have positions clearly indicating that they have no contamination from background or foreground objects or structures. 

(2) For comparison purposes, we have analyzed archival {\it Chandra} data of the three galaxies. 13 of these 16 {\it XMM-Newton} sources have also been detected by {\it Chandra}. Their source positions, in general,  show a clear offset  $\sim$ 2$\arcsec$  from the center of {\it XMM-Newton} error circle (15$\arcsec$). 
 Only two of these 13 sources have been examined spectrally due to sufficient statistics provide by {\it Chandra} observations. 


(3) Four of these 16 sources  are ULX candidates  owing to their spectral characteristics and high luminosities. Two of these sources are in NGC 4258 (XMM-2 and XMM-3) and they are  best fitted with an absorbed DISKBB model with a hot inner disk temperature ({\it T$_{in}$} $\sim$  1.43 keV and 1.28 keV respectively).  These values are within  the range  given for Galactic black-hole XRBs in the high-state emission (McClintock \& Remillard 2006). On the other hand, 2  ULXs (XMM-2 in NGC 4736 and XMM-2 in NGC 4258) show significant transient behavior.  The statistically significant improvement over a one-component fit is found in the PL+DISKBB model  fit  for  the ULX in NGC 4736. The disk temperature is also cool (0.75 keV) in this case. Similar disk temperatures for ULX sources in M101 (Jenkins et al. 2004) and in the interacting pair galaxies NGC 4485/4490 (Gladstone \& Roberts, 2009) were already found. Based on the argument that  black hole masses scale inversely with accretion disc temperatures ({\it T$_{in}$} $\sim$ M$^{-1/4}$),  this was interpreted as evidence of intermediate  mass black holes (IMBHs), with masses in the range of 10$^{2}$-10$^{4}$ M$_{\odot}$ for some ULX sources (Colbert \& Mushotzky 1999). On the other hand, assuming spherical accretion onto  a central black-hole in a binary system at the Eddington limit as  suggested by  Makishima et al (2000), our point source luminosities imply black hole masses in the stellar range of (2-15) M$_{\odot}$. This is consistent with Galactic stellar-mass black holes which were found to lie in  the range of (2-23) M$_{\odot}$  (McClintock $\&$  Remillard 2004).  Most probably, these sources belong to  the extreme end of  the XRB population in their galaxies. Many similar luminous point sources in nearby galaxies are also reported (Jenkins et al. 2004;  Gladstone $\&$ Roberts 2009).

(4) Of the remaining 12 sources, 8 are best fitted with absorbed PL spectra ($\Gamma$ =  1.4 - 2.8). They generally have luminosities around  the Eddington limit ($\geq$ 10$^{38}$). Photon indices of 7 of these sources are in agreement with low/hard state black-hole XRBs (Tanaka 2001). Remaining  2 sources (XMM-10 and XMM-16 in NGC 4258) have steeper power-law slopes ($\Gamma$ $>$ 2) representing strong Comptonized thermal emission from a black-hole accretion disc (Jenkins et al. 2004,  McClintock  $\&$ Remillard 2004). Based on X-ray data these sources are considered to be black-hole XRBs.  

(5)  The locations of the point sources could also provide clues to the nature of the X-ray emission; however, at times they could also be misleading.  Positions of our sources suggest that almost all are associated with their host galaxies except possibly one. As explained above, this latter source (XMM-6 in NGC 4258) could be a background AGN due to its spectral shape (PL with $\Gamma$ $\sim$ 1.9  which  is typical of AGNs) and its coincidence with a stellar object (in corresponding DSS image), warranting further investigation.

(6) A noticeable overall result of the spectral analysis of the sources from these three galaxies, is that X-ray sources from different  classes can be distinguished. One source (XMM-5 of NGC 4395)  is distinctly different from rest. A closer inspection by our optical follow up observation revealed that, it was indeed different type; an SNR with a best spectral fit to a VPSHOCK, typical of such objects (Sonbas et al. 2010). 

(7) We have identified only one SSS  with typical low luminosity values around 5$\times$10$^{37}$ erg s$^{-1}$ in one of these galaxies (XMM-23 in NGC 4395). This class of sources generally are known to be a white dwarf accreting at high rates resulting in nuclear burning on the surface  with \textit{kT} $\sim$ (10-100) eV (Di Stefano et al. 2004; Stiele et al. 2010). Our source also confirms these limits.

(8) About half of presently analyzed sources show statistically significant short and long- term variabilities. This can be taken as evidence for the underlying binary nature of these sources.

In more general terms, we can conclude that X-ray astronomy as reached at a stage to differentiate several source types in the galaxies beyond our local group. With increasing statistics, the role of such sources in the evolution of stars and such galaxies will be clearer.

\acknowledgments

Aysun Akyuz acknowledges support from the EU FP6 Transfer of Knowledge Project "Astrophysics of Neutron Stars" (MKTD-CT-2006-042722). We thank Frank Haberl and Nazim Aksaker for their very valuable help. We also thank the TUBITAK National Observatory (TUG)  for their support for observing times and equipment.

\clearpage



\begin{figure*}
\centering
 (a)
{
\label{fig:sub:a}
  \includegraphics[width=6.0cm]{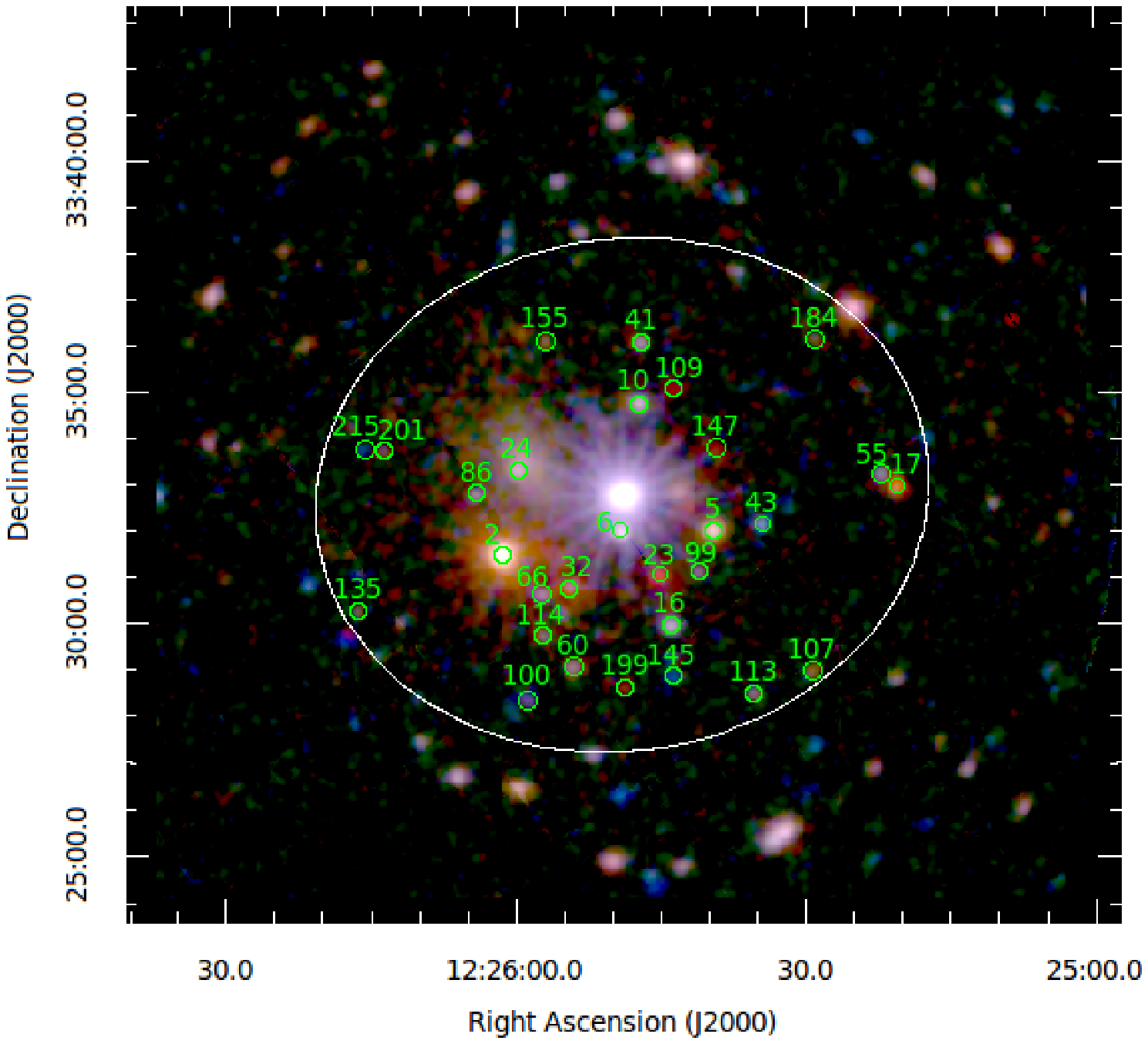}
}
\hspace{1.5cm}
 (b)
{
\label{fig:sub:b}
\includegraphics[width=5.5cm]{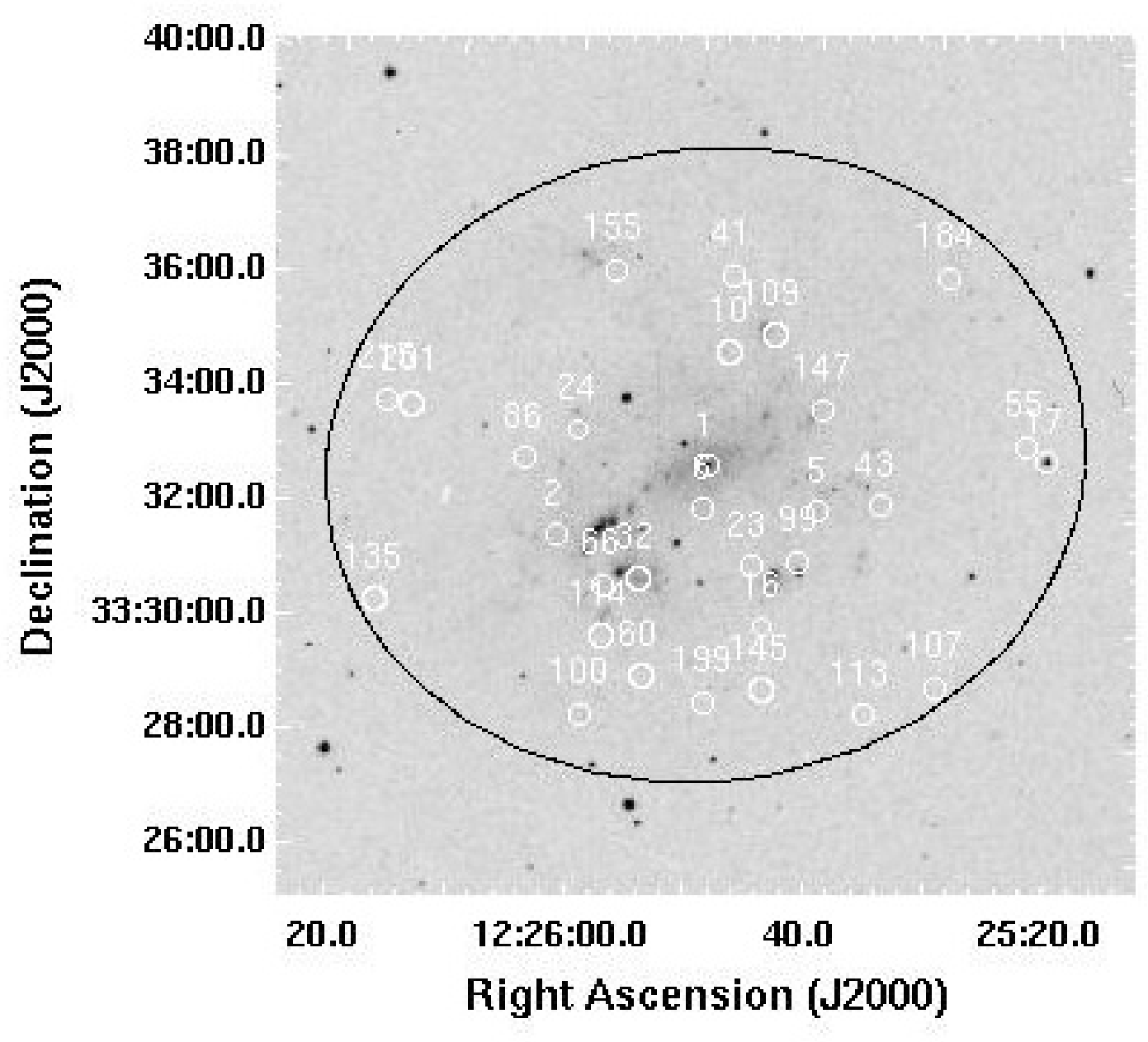}
    \caption{ In (a), {\it XMM-Newton} RGB image of NGC 4395. Colors stand for red (0.2-1) keV, green (1-2) keV and blue (2-12) keV. The {\it D$_{25}$} ellipse is also shown for comparison. In (b), {\it XMM-Newton} source positions are overlaid on a DSS1 optical blue image  (circles are not representing positional error radii). The detected point sources inside the {\it D$_{25}$} ellipse are specified as off-nuclear galactic objects.}
    \label{<ngc4395>}
}
\label{fig:sub}
\end{figure*}


\begin{figure*}
\centering
 (a)
{
\label{fig:sub:a}
  \includegraphics[width=6.0cm]{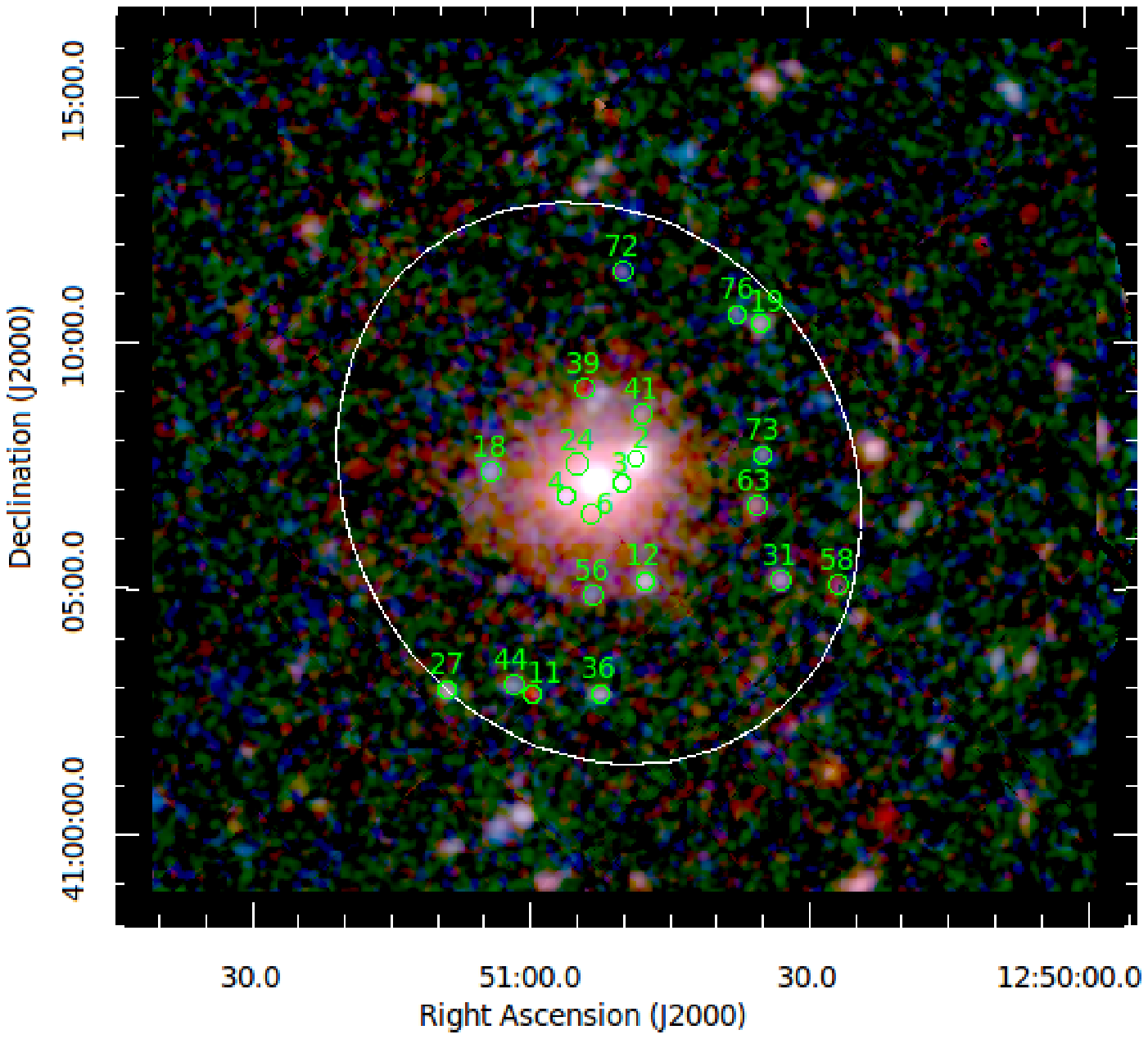}
}
\hspace{1.5cm}
 (b)
{
\label{fig:sub:b}
\includegraphics[width=5.5cm]{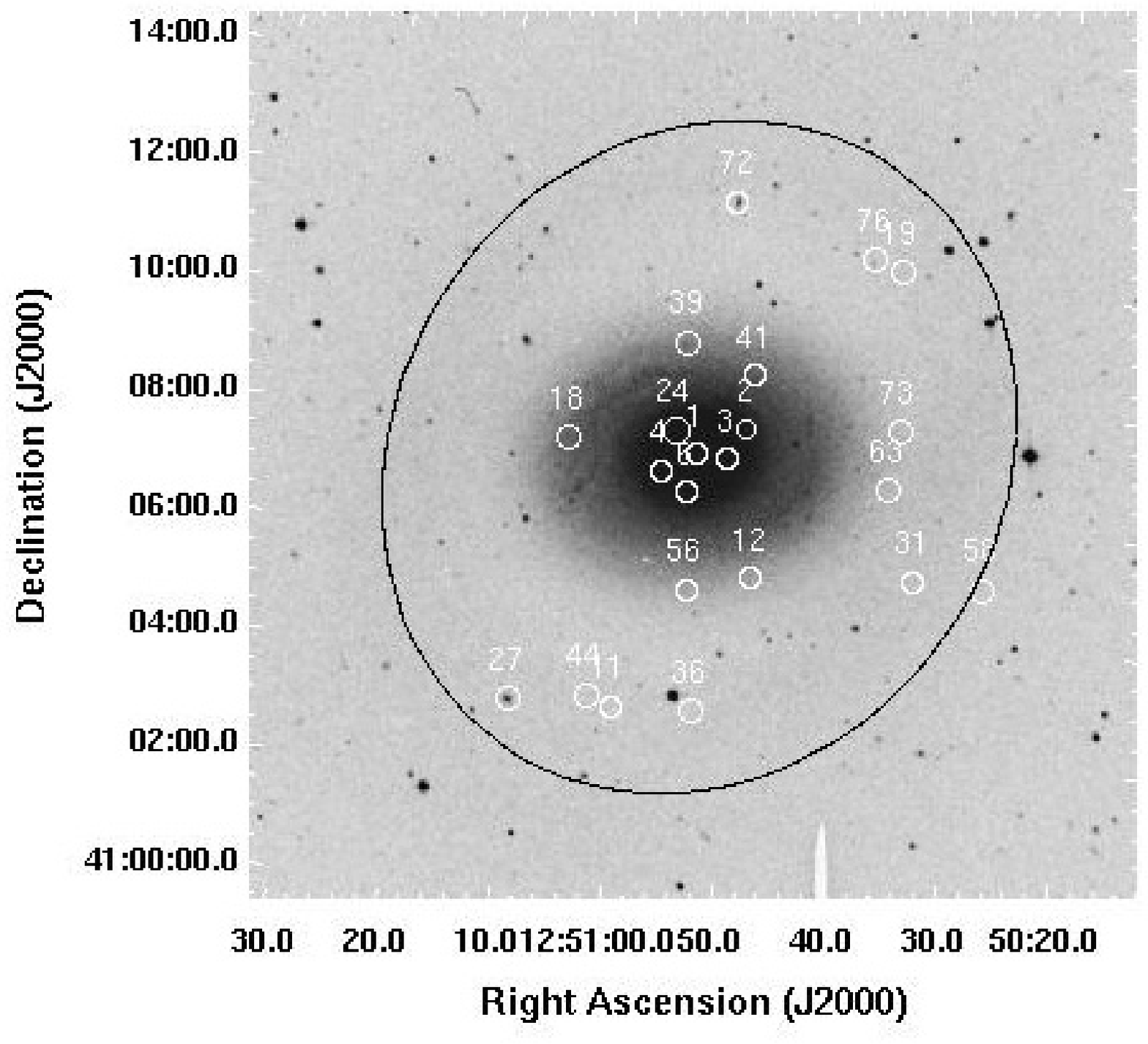}
    \caption{{\it XMM-Newton} RGB image of NGC 4736. Definitions are the same as in Figure 1.}
    \label{<ngc4736>}
} 
\end{figure*}

\begin{figure*}
\centering
\hspace{0.5cm}
 (a)
{
\label{fig:sub:a}
  \includegraphics[width=6.0cm, height=6cm]{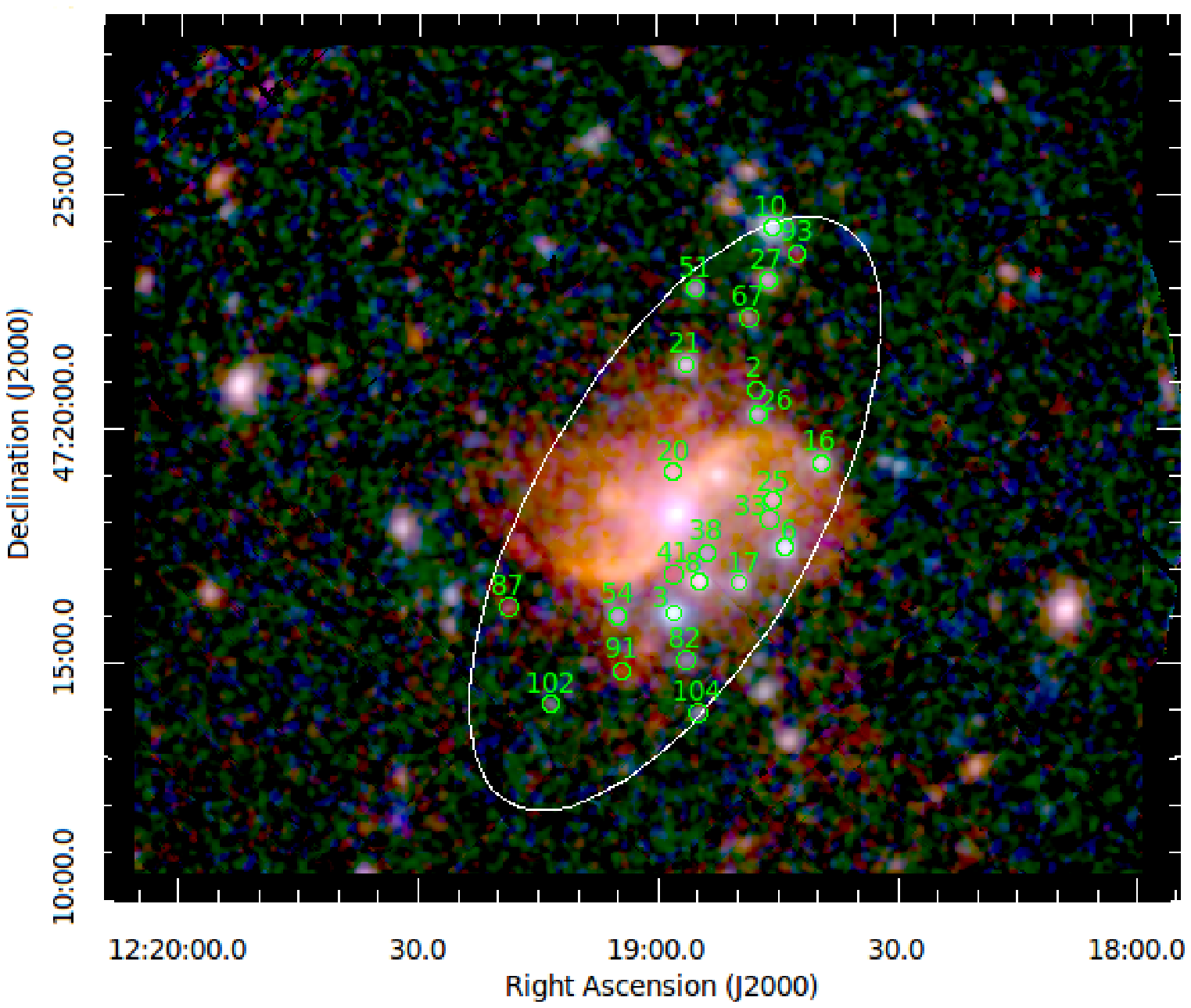}
}
\hspace{0.25cm}
 (b)
{
\label{fig:sub:b}
\hspace{1.0cm}
\includegraphics[width=6.6cm]{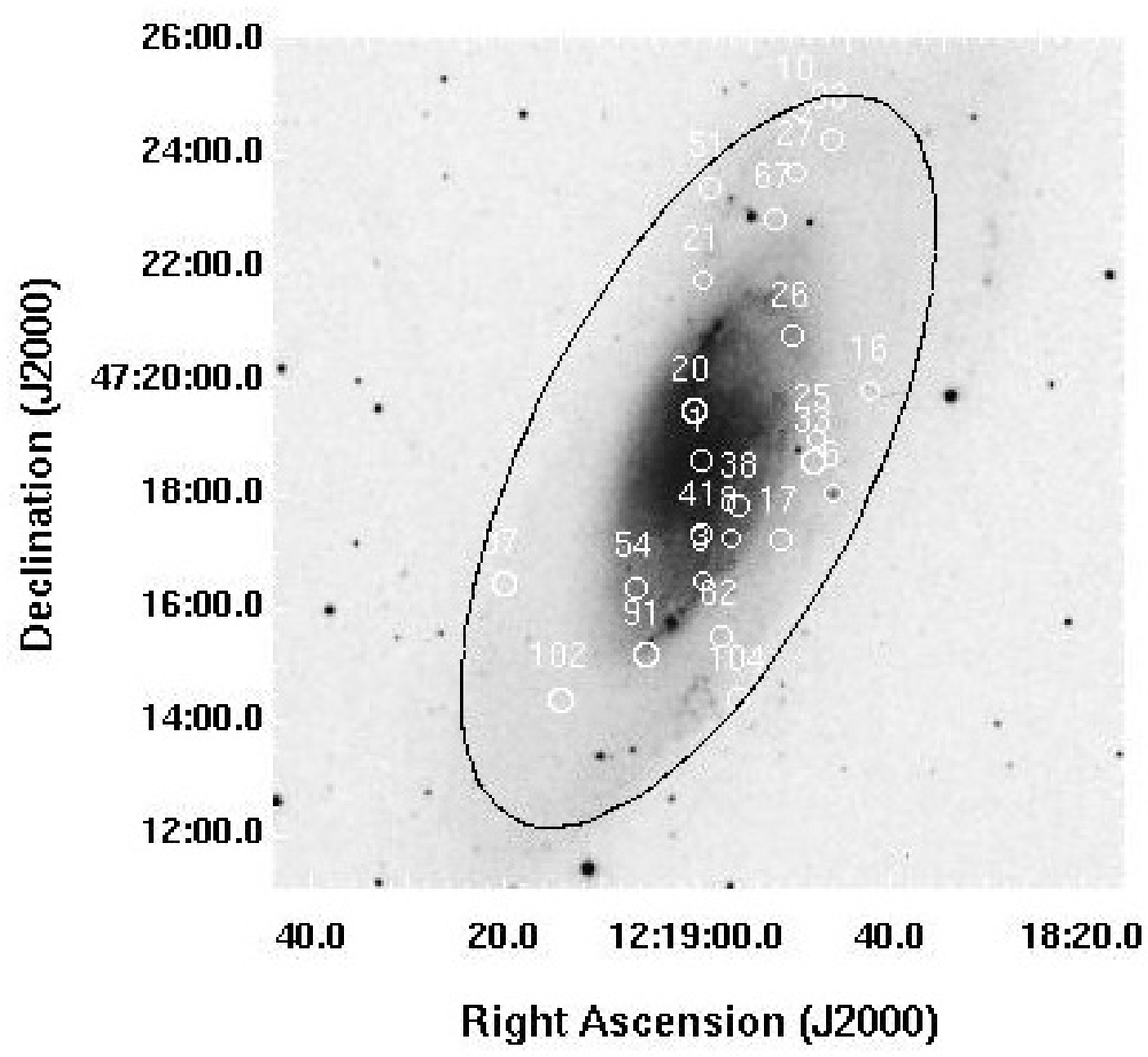}
    \caption{{\it XMM-Newton} RGB image of NGC 4258. Definitions are the same as in Figure 1}
    \label{<ngc4258>}
}
\end{figure*}


\clearpage


\begin{figure*}
\centering

\includegraphics[width=17cm, angle=0]{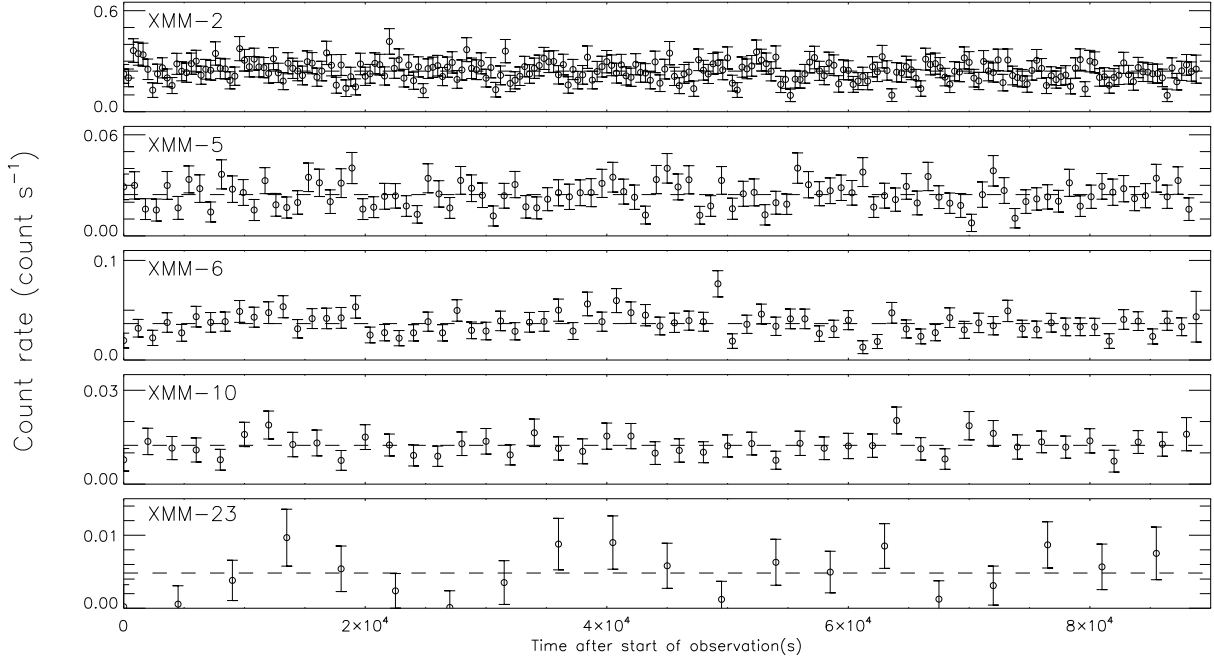}
\caption{{\it XMM-Newton} short-term light curves of NGC 4395 point sources. These light curves were obtained from the counts using only EPIC-pn camera.  A dashed line shows the mean count rate and error bars correspond to 1$\sigma$ deviations assuming Gaussian statistics.}

\end{figure*}


\begin{figure*}
\centering
  \includegraphics[width=17cm, angle=0]{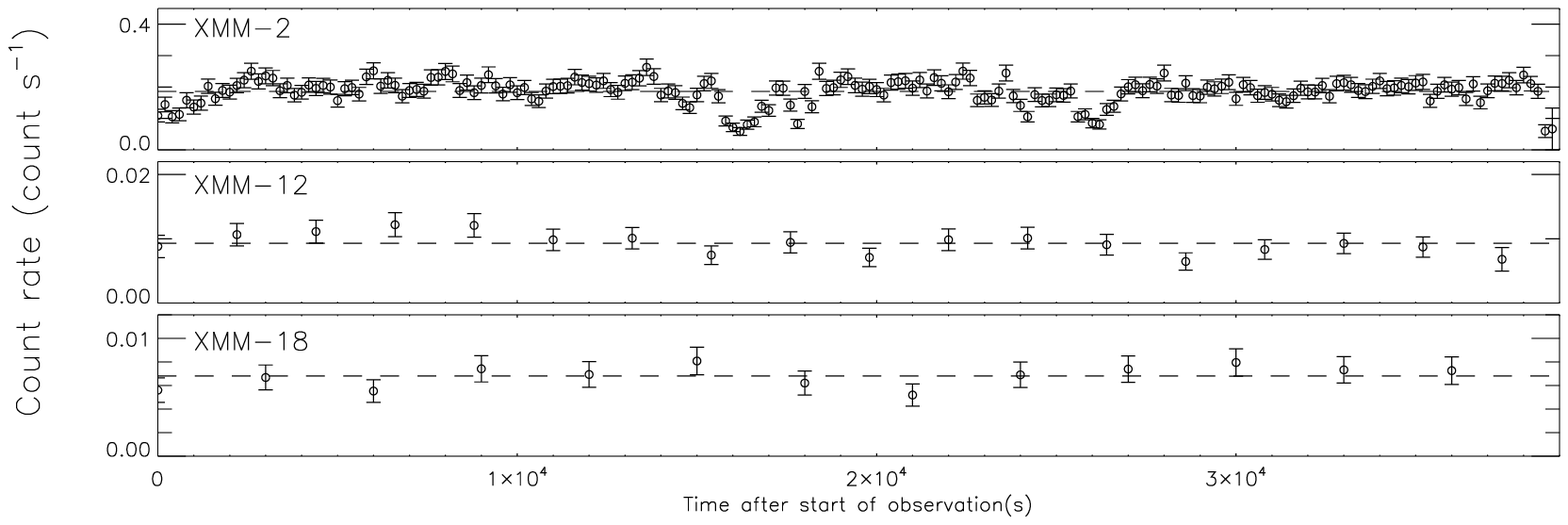}
    \caption{{\it XMM-Newton} short-term light curves of NGC 4736 point sources. These light curves were  obtained by summing the counts from the EPIC-pn and MOS cameras. A dashed line shows the mean count rate and error bars correspond to 1$\sigma$ deviations assuming Gaussian statistics.}
    \label{<Your label>}
\end{figure*}

\begin{figure*}
\centering
  \includegraphics[width=17cm, angle=0]{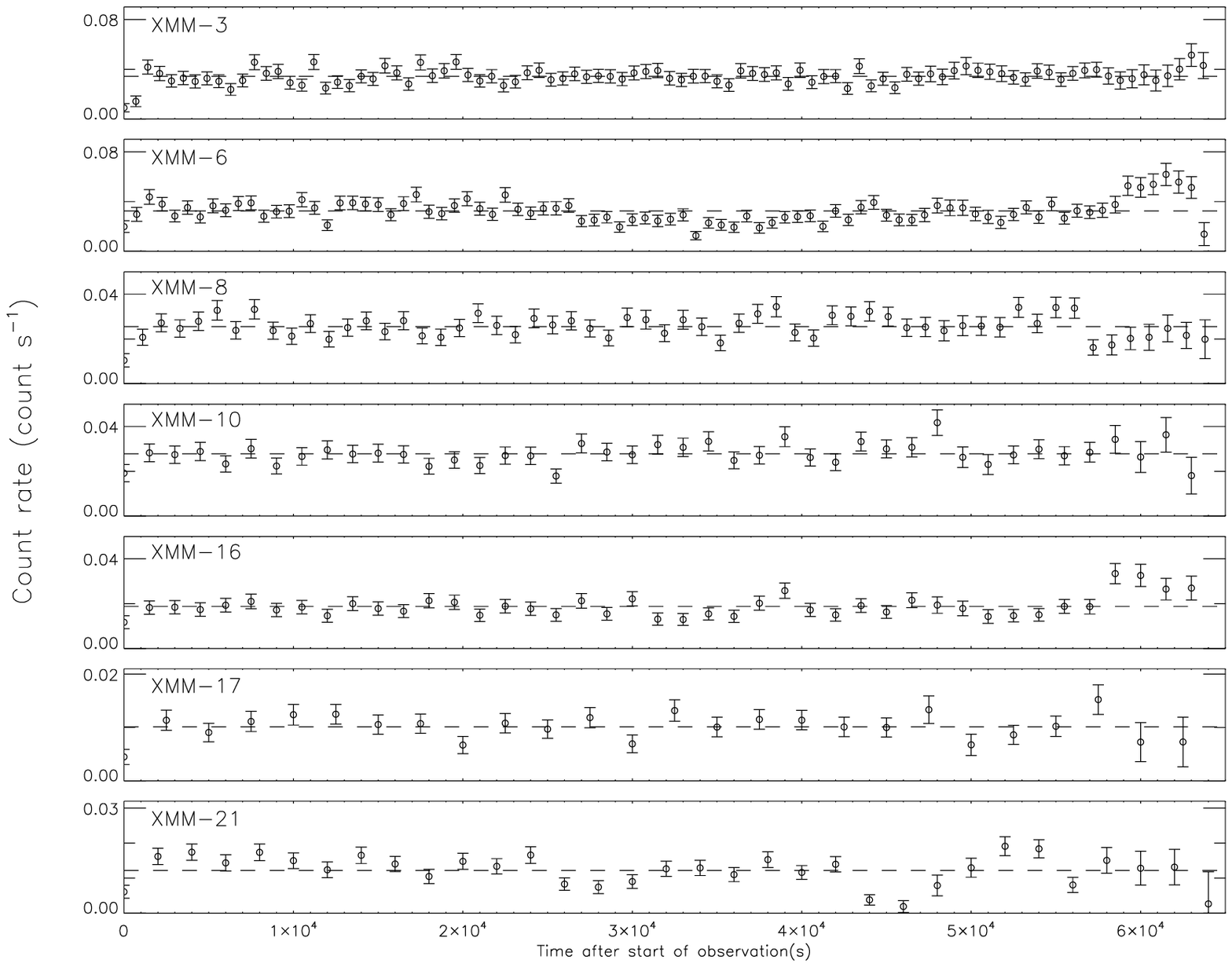}
    \caption{{\it XMM-Newton} short-term light curves of NGC 4258 point sources. These light curves were  obtained by summing the counts from the EPIC-pn and MOS cameras. A dashed line shows the mean count rate and error bars correspond to 1$\sigma$ deviations assuming Gaussian statistics.}
    \label{<Your label>}
\end{figure*}


\begin{figure*}
\centering
 (a)
{
\label{fig:sub:a}
\includegraphics[width=16cm, height=5.5cm, angle=0]{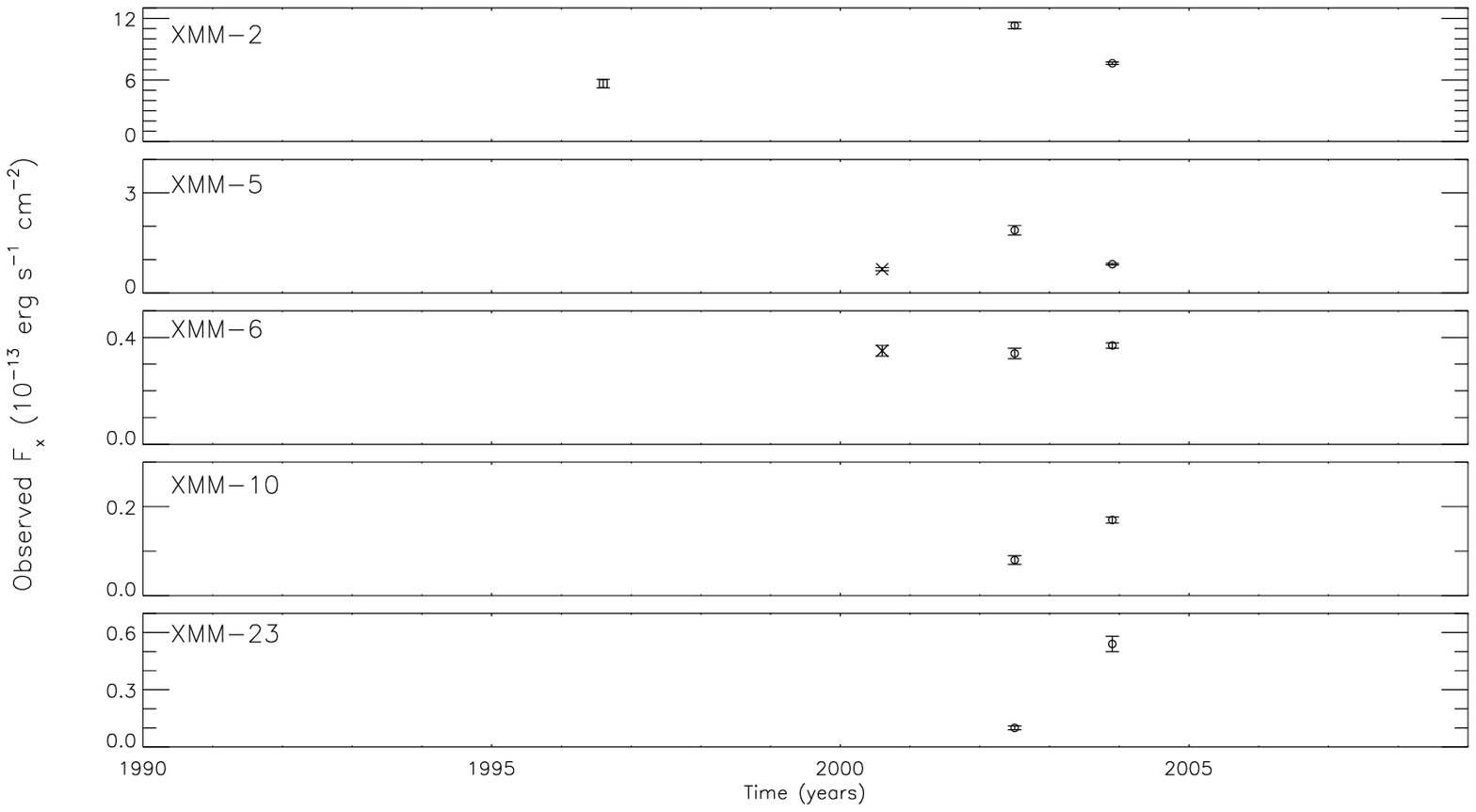}
}
\hspace{1.1cm}
 (b)
{
\label{fig:sub:b}
\includegraphics[width=16cm, height=3.5cm, angle=0]{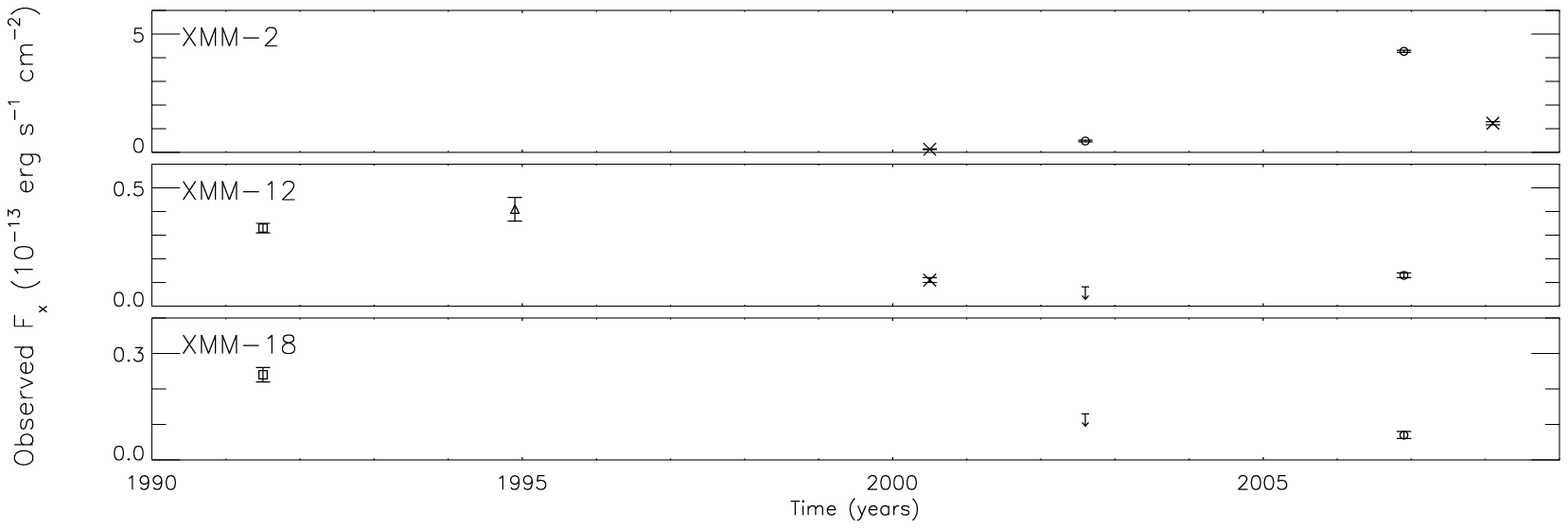}
}
\hspace{1.1cm}
 (c)
{
\label{fig:sub:c}
\includegraphics[width=16cm, height=8.5cm, angle=0]{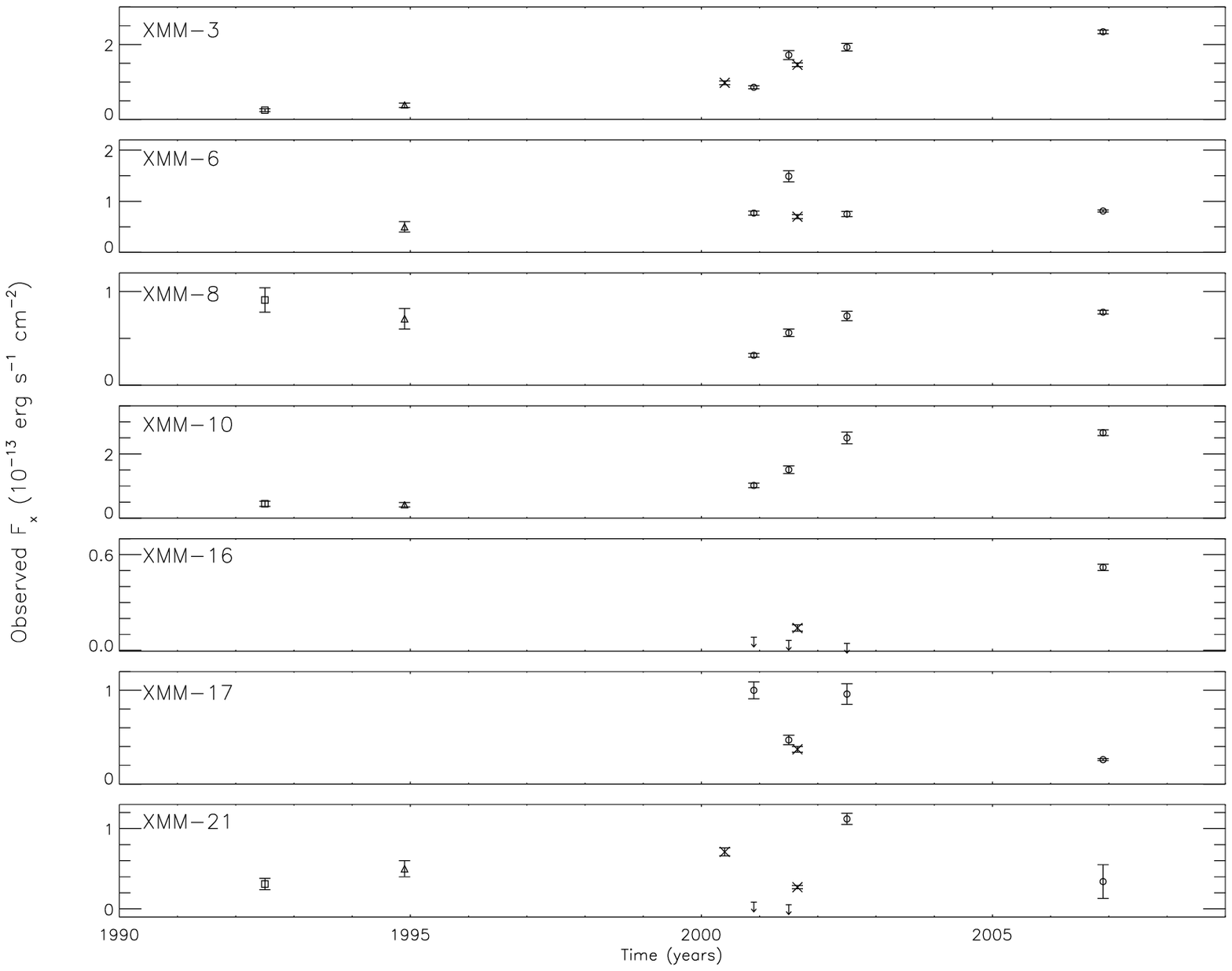}
}
\caption{XMM-Newton long-term light curves of NGC 4395, NGC 4736 and NGC 4258 as (a), (b) and (c) respectively. Data points were taken from {\it XMM-Newton} (circles), {\it ROSAT PSPC} (squares), {\it ROSAT HRI} (triangles), {\it Chandra} (stars) flux measurements have been converted to  (0.5 - 2) keV flux as explained in the text. Flux upper limits are shown with downward arrows.}

\label{fig:sub}
\end{figure*}
\begin{figure*}
\centering
 (a)
{
\label{fig:sub:a}
\includegraphics[width=17cm, angle=0]{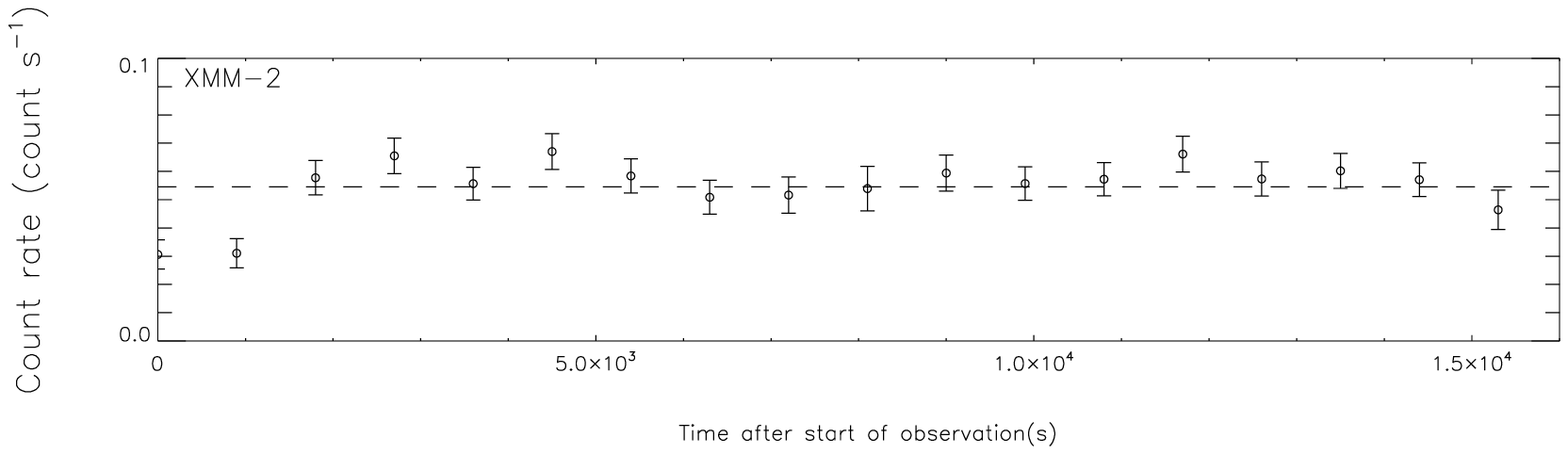}
}
\hspace{1.5cm}
 (b)
{
\label{fig:sub:b}
\includegraphics[width=17cm, angle=0]{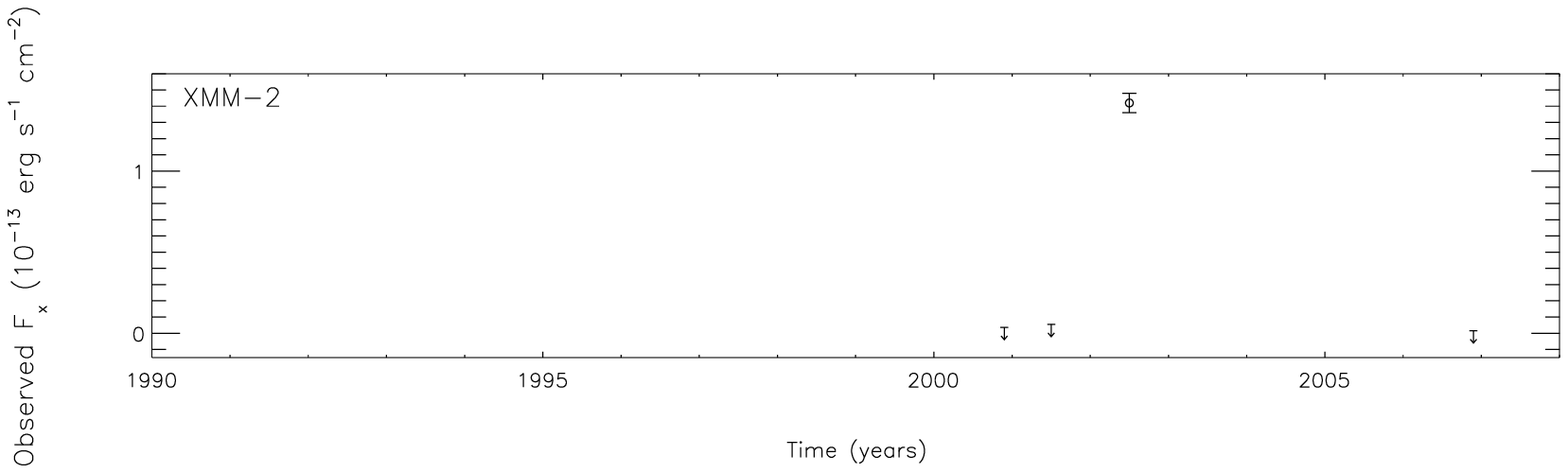}
}

\caption{(a) short-term  (b) long-term light curves of XMM-2 in NGC 4258. Properties and possible nature of this transient source are described in the text.}

\label{fig:sub}
\end{figure*}

\begin{figure*}
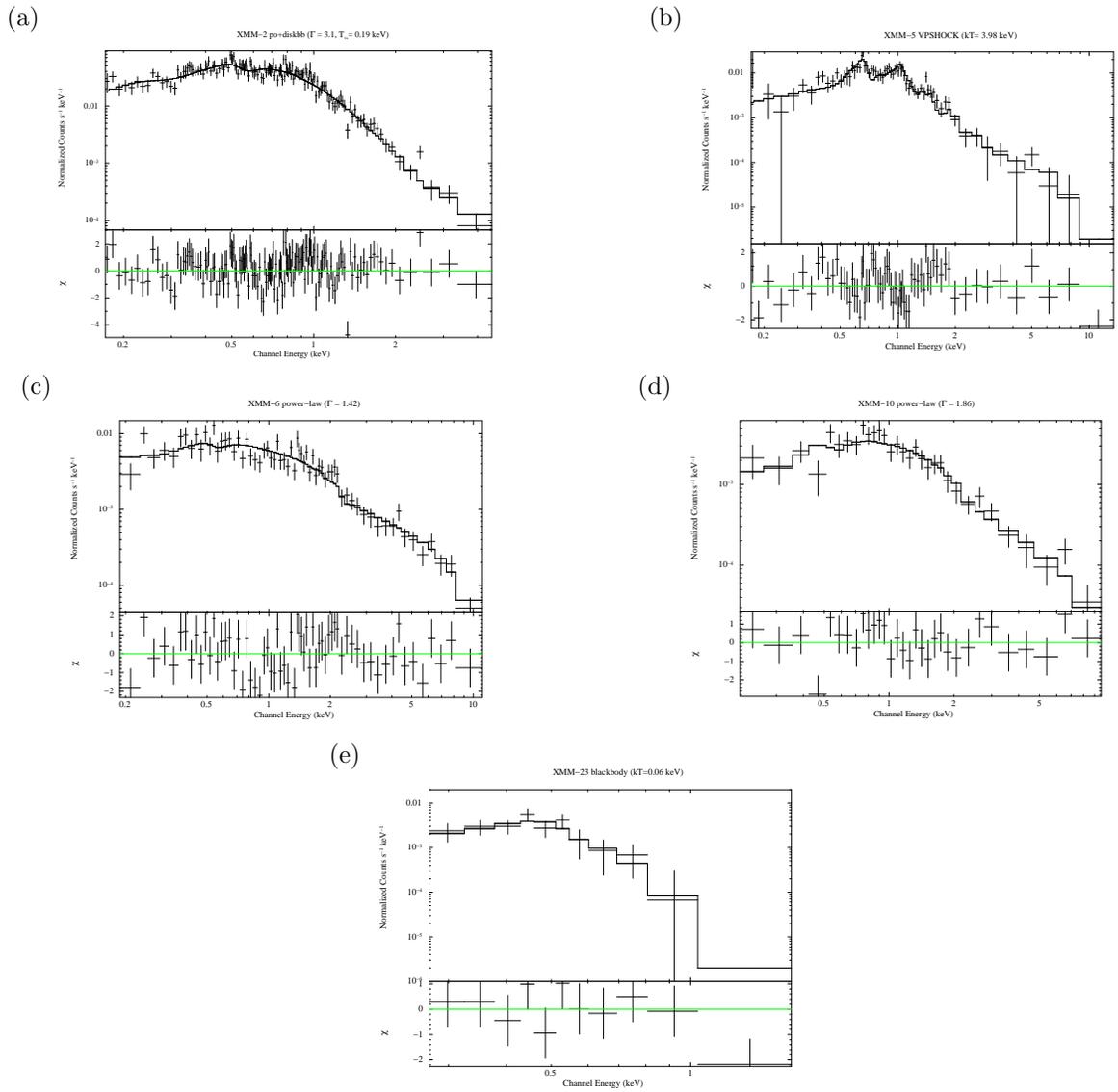

\centering
 (a)
{
\label{fig:sub:a}
\includegraphics[width=4.8cm, angle=-90]{Fig10_ngc4395_2_pobbody.ps}
}
\hspace{1cm}
 (b)
{
\label{fig:sub:b}
\includegraphics[width=4.8cm, angle=-90]{Fig10_ngc4395_5_vpshock.ps}
}
\hspace{1cm}
 (c)
{
\label{fig:sub:c}
\includegraphics[width=4.8cm, angle=-90]{Fig10_ngc4395m_6_lessbck_po.ps}
}
\hspace{1cm}
 (d)
{
\label{fig:sub:d}
\includegraphics[width=4.8cm, angle=-90]{Fig10_ngc4395_10_po.ps}
{
\hspace{1cm}
 (e)
}
\label{fig:sub:e}
\includegraphics[width=4.8cm, angle=-90]{Fig10_ngc4395_23_lessbck_bbody.ps}
\caption{The best fitting model, spectra  (upper panels), and the
residuals between the data and the model in standard deviations
(lower panels) of the sources in NGC 4395. Since the  MOS data has poor statistics, only the EPIC-pn spectra have been obtained  for these sources.}
}
\label{fig:sub}
\end{figure*}

\begin{figure*}
\centering
  \includegraphics[height=8cm, width=9cm]{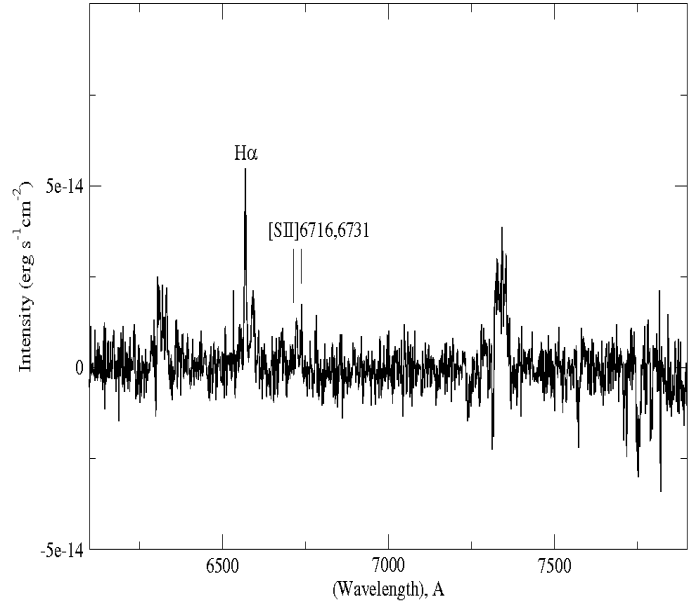}
   \caption{Optical spectrum of XMM5 in NGC 4395 obtained with the TFOSC instrument on RTT150 at TUG, Antalya, Turkey}
   \label{xmm5}
\end{figure*}
\begin{figure*}
\centering
 (a)
{
\label{fig:sub:a}
\includegraphics[width=4.8cm, angle=-90]{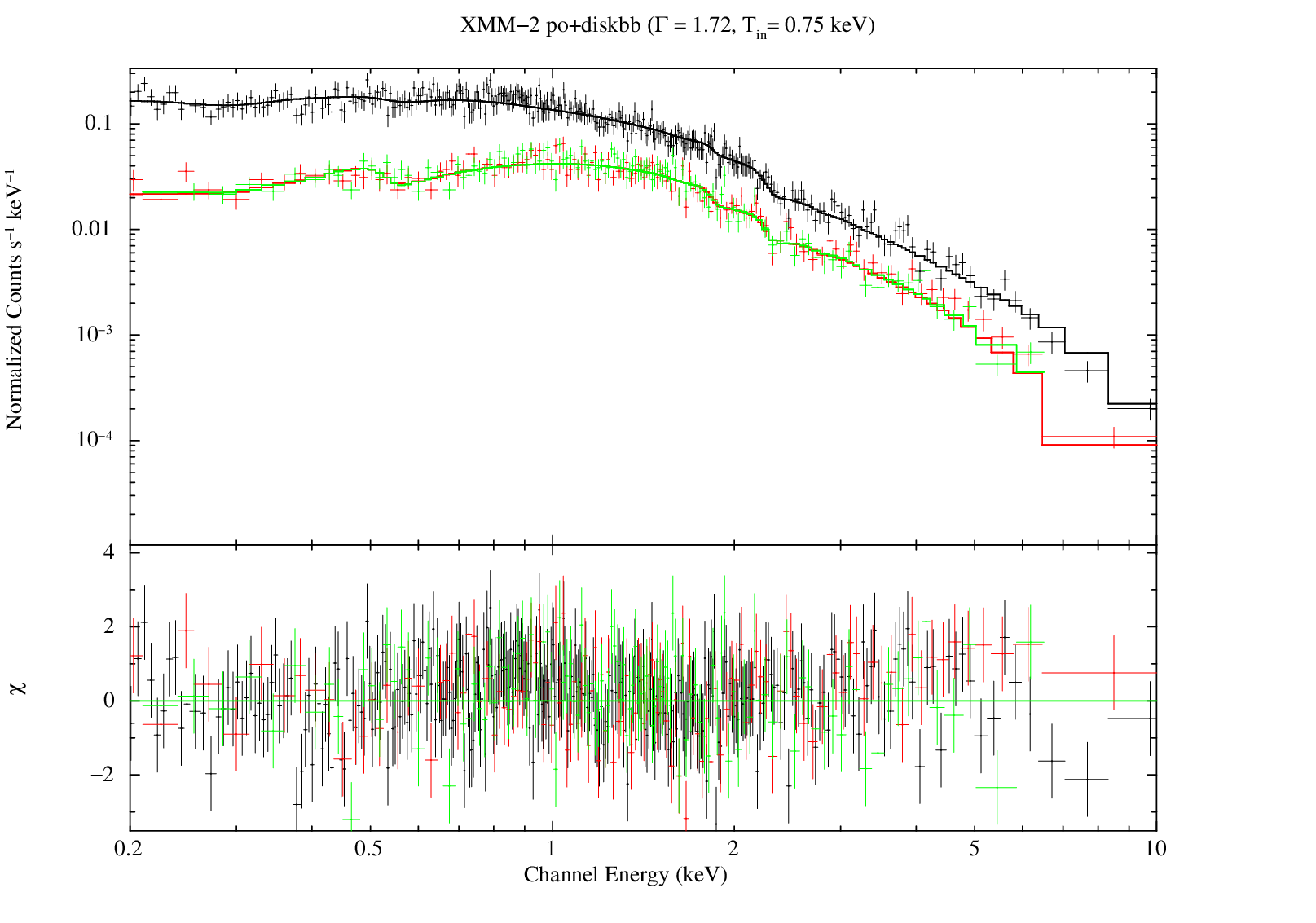}
}
\hspace{0.8cm}
 (b)
{
\label{fig:sub:b}
\includegraphics[width=4.8cm, angle=-90]{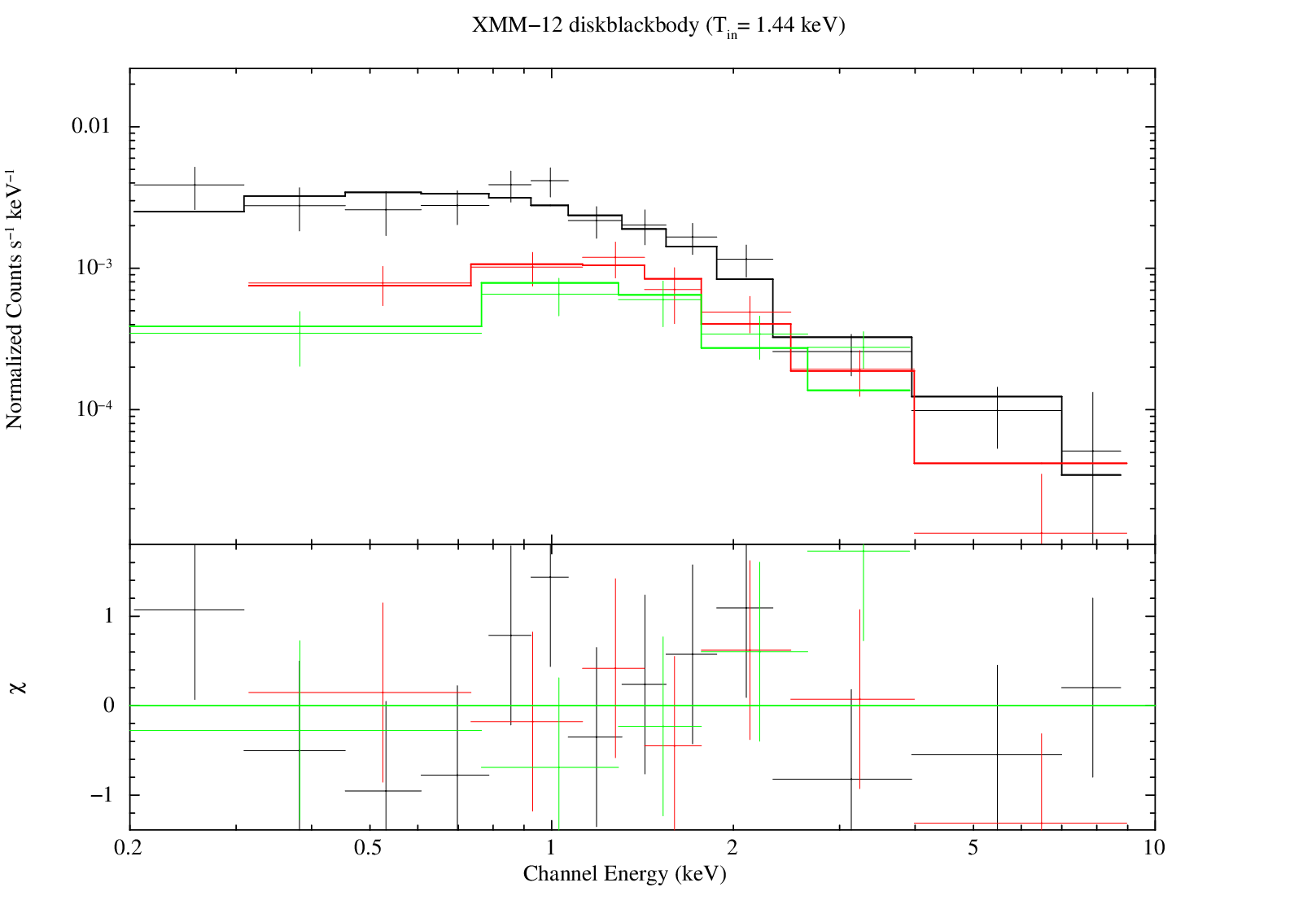}
}
\hspace{0.8cm}
 (c)
{
\label{fig:sub:c}
\includegraphics[width=4.8cm, angle=-90]{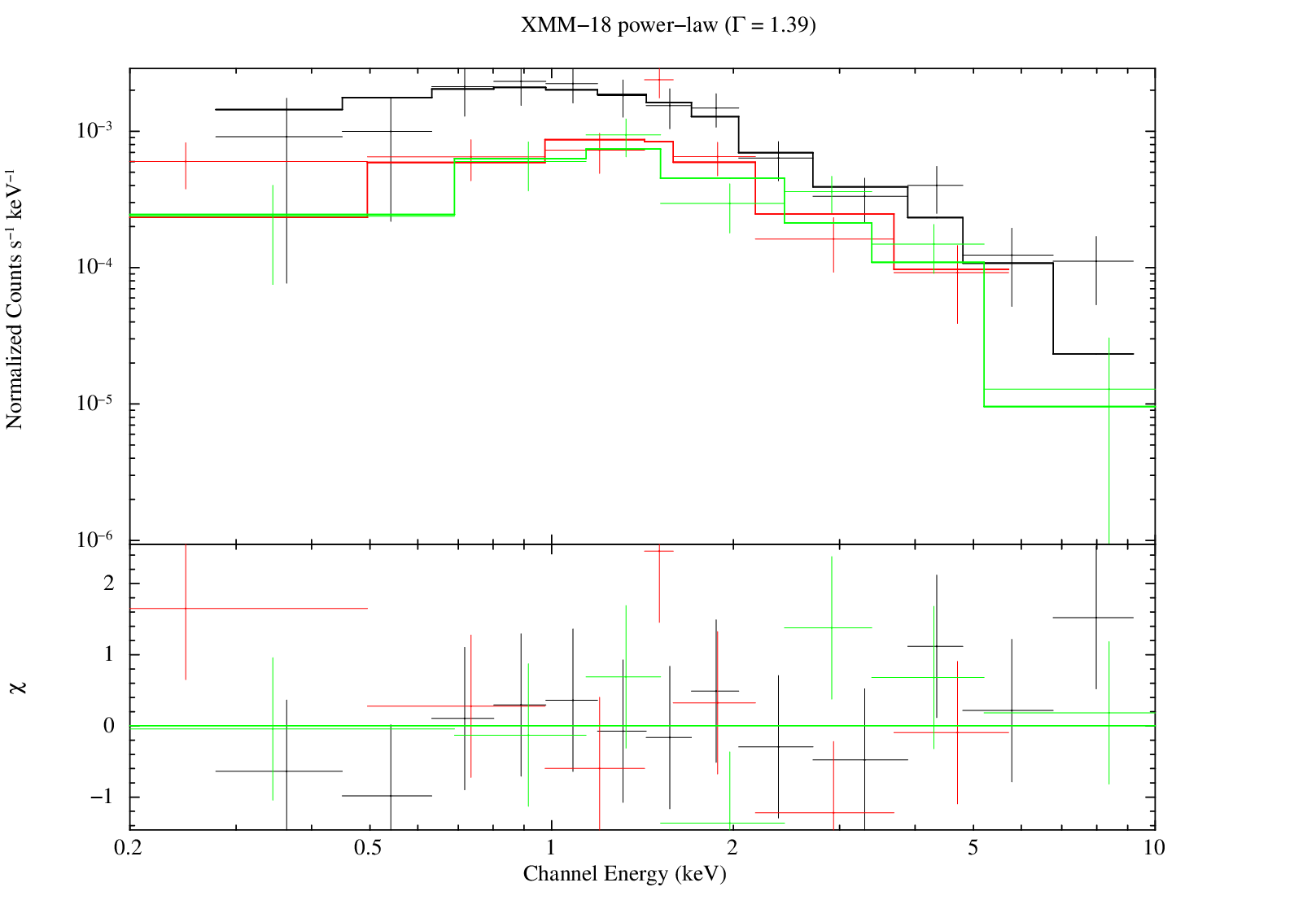}
\caption{The best fitting model, spectra  (upper panels), and  the
residuals between the data and the model in standard deviations
(lower panels) of the sources in NGC 4736. The EPIC-pn data points and best fitting model are shown in black; data from MOS1 and MOS2  are shown in green and red, respectively.}
}
\label{fig:sub}
\end{figure*}

\begin{figure*}
\centering
 (a)
{
\label{fig:sub:a}
\includegraphics[width=4.8cm, angle=-90]{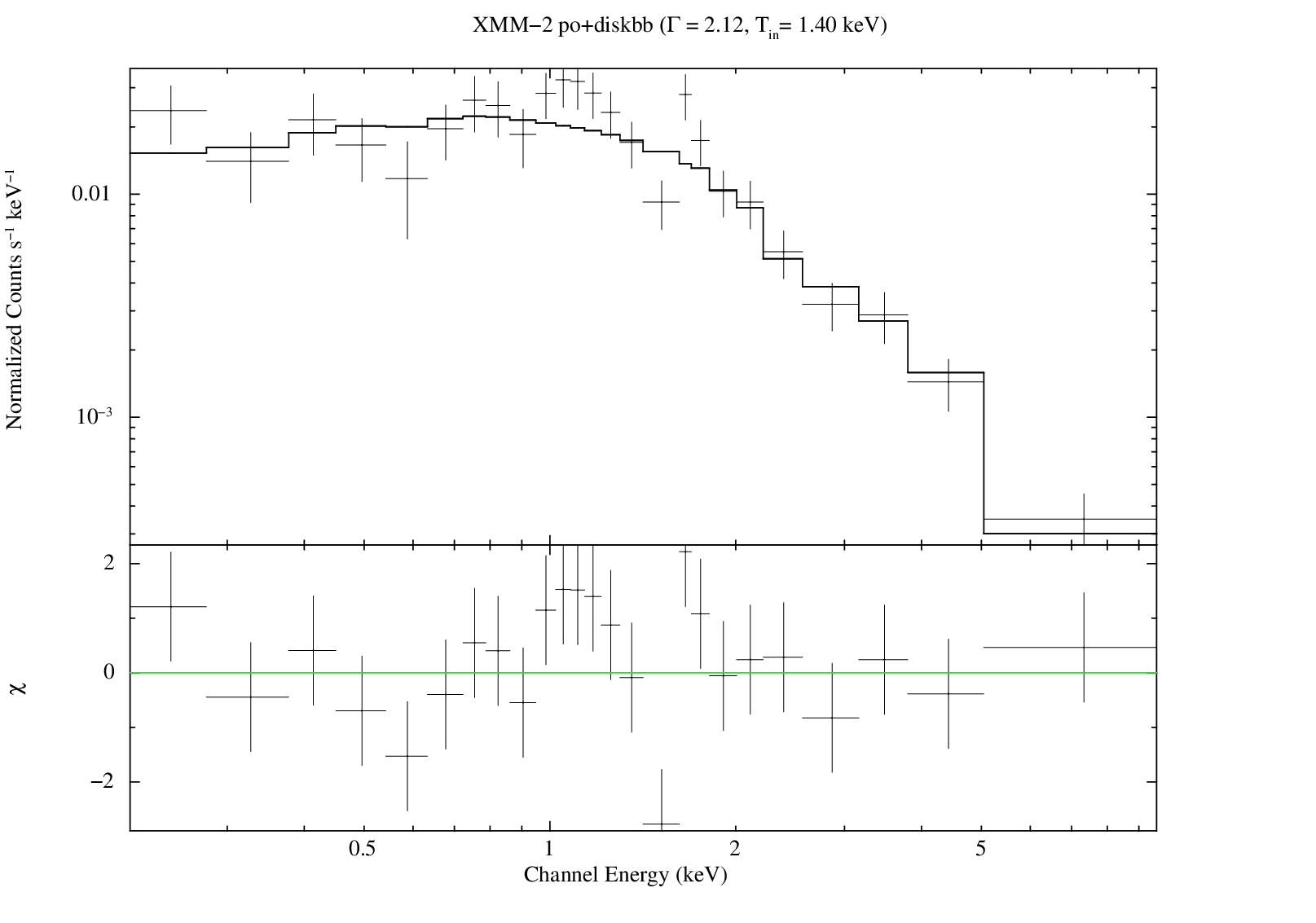}
}
\hspace{0.8cm}
 (b)
{
\label{fig:sub:b}
\includegraphics[width=4.8cm, angle=-90]{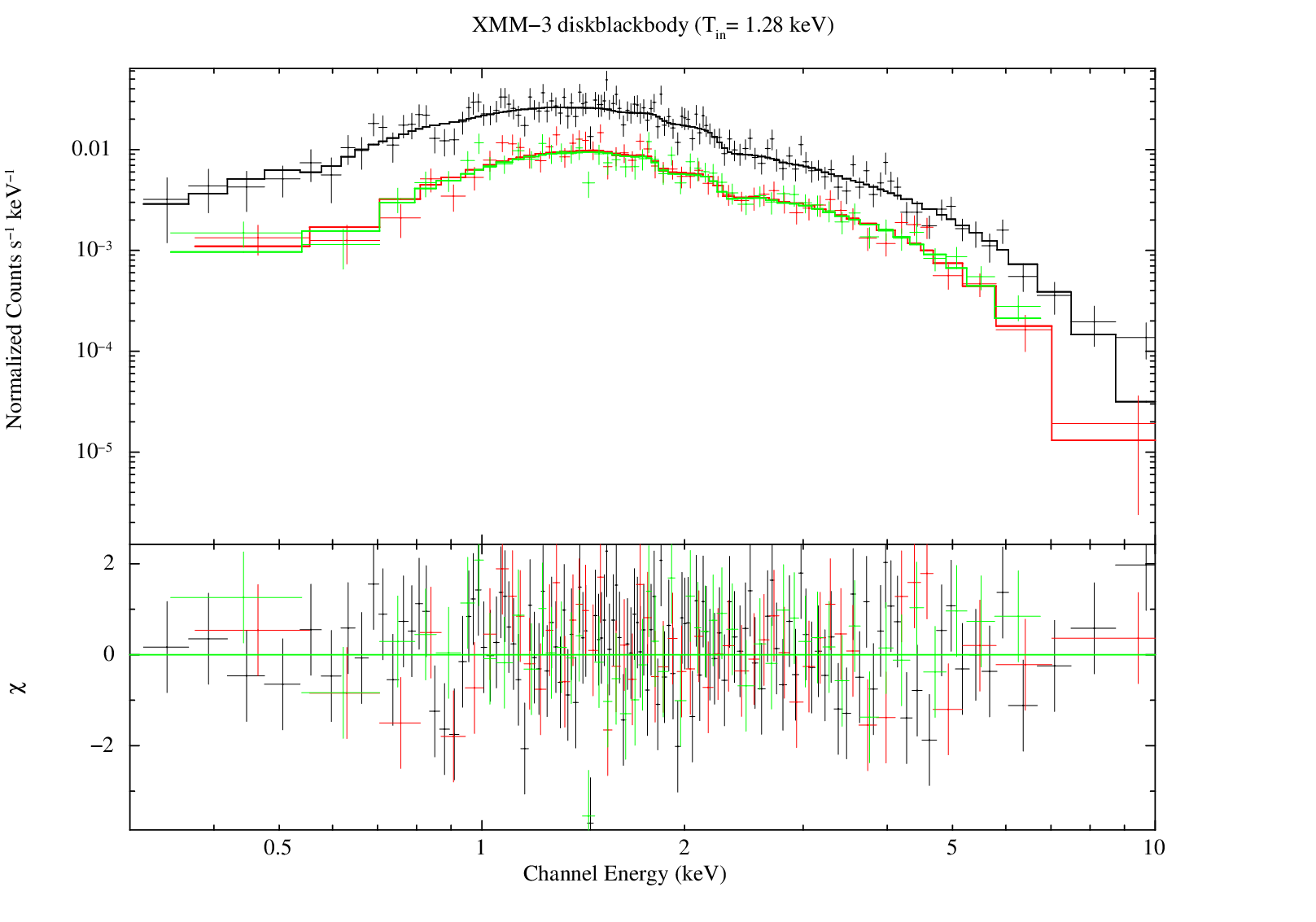}
}
\hspace{0.8cm}
 (c)
{
\label{fig:sub:c}
\includegraphics[width=4.8cm, angle=-90]{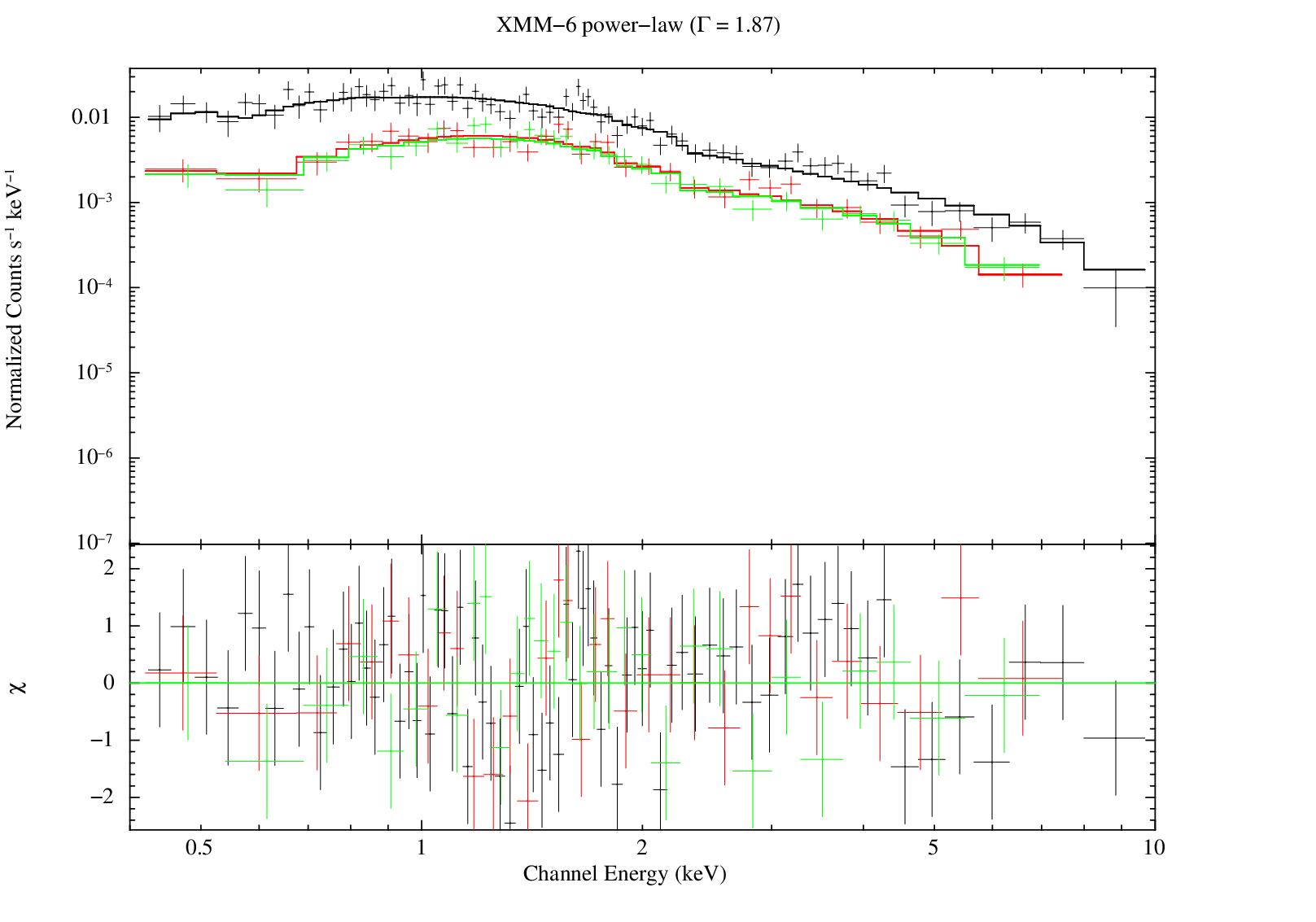}
}
\hspace{0.8cm}
 (d)
{
\label{fig:sub:d}
\includegraphics[width=4.8cm, angle=-90]{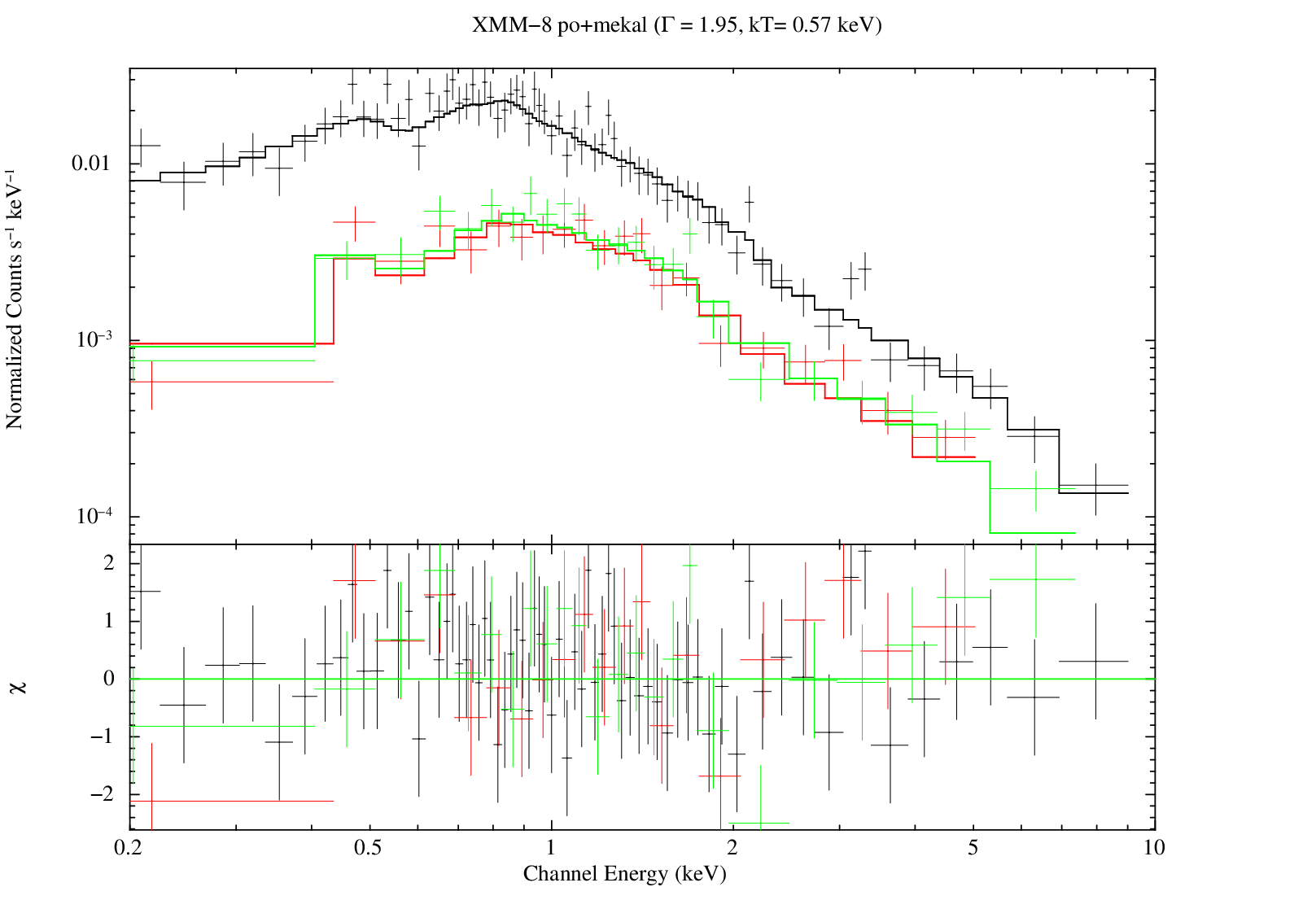}
}
\hspace{0.8cm}
 (e)
{
\label{fig:sub:e}
\includegraphics[width=4.8cm, angle=-90]{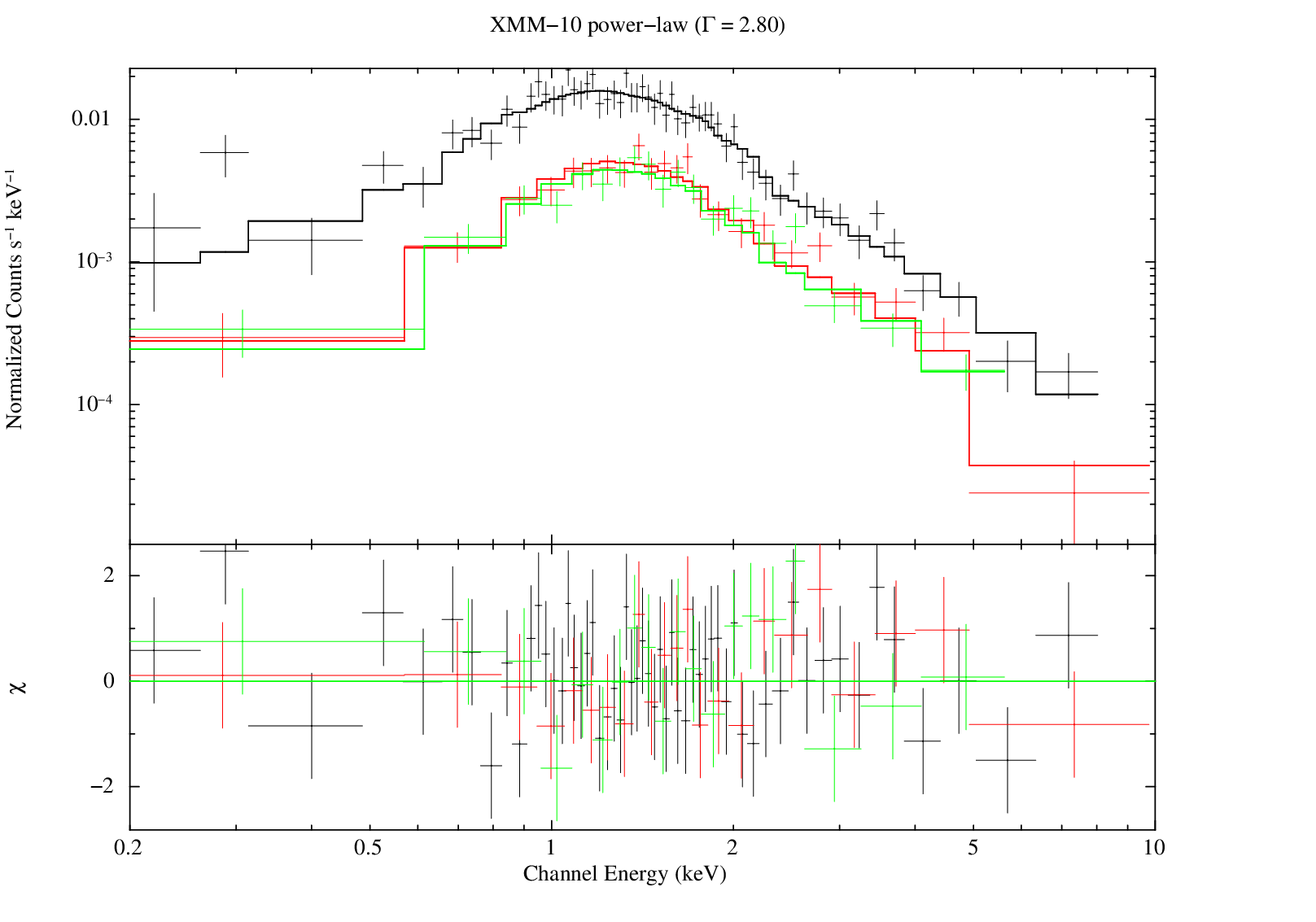}
}
\hspace{0.8cm}
 (f)
{
\label{fig:sub:f}
\includegraphics[width=4.8cm, angle=-90]{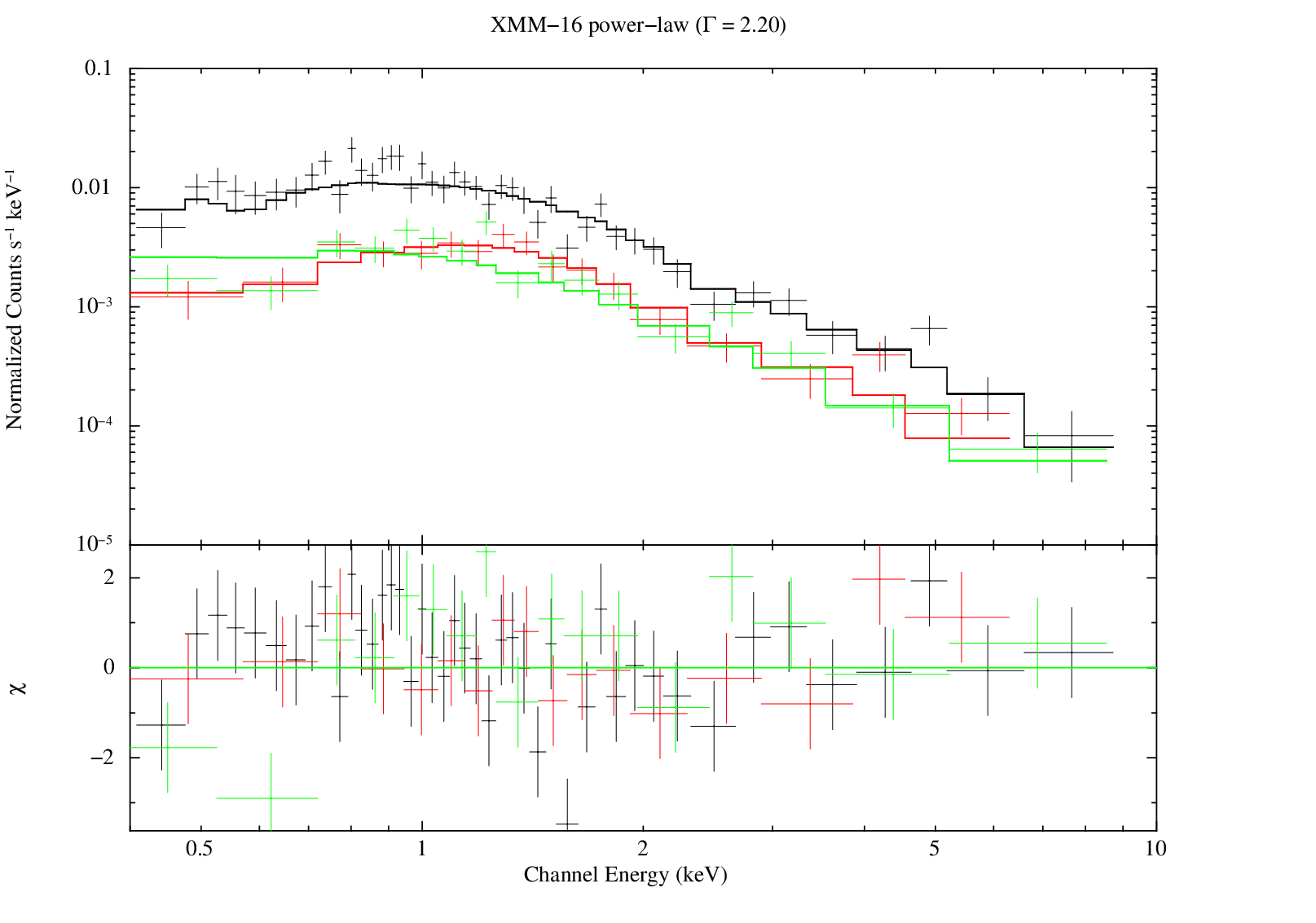}
}
\hspace{0.8cm}
 (g)
{
\label{fig:sub:g}
\includegraphics[width=4.8cm, angle=-90]{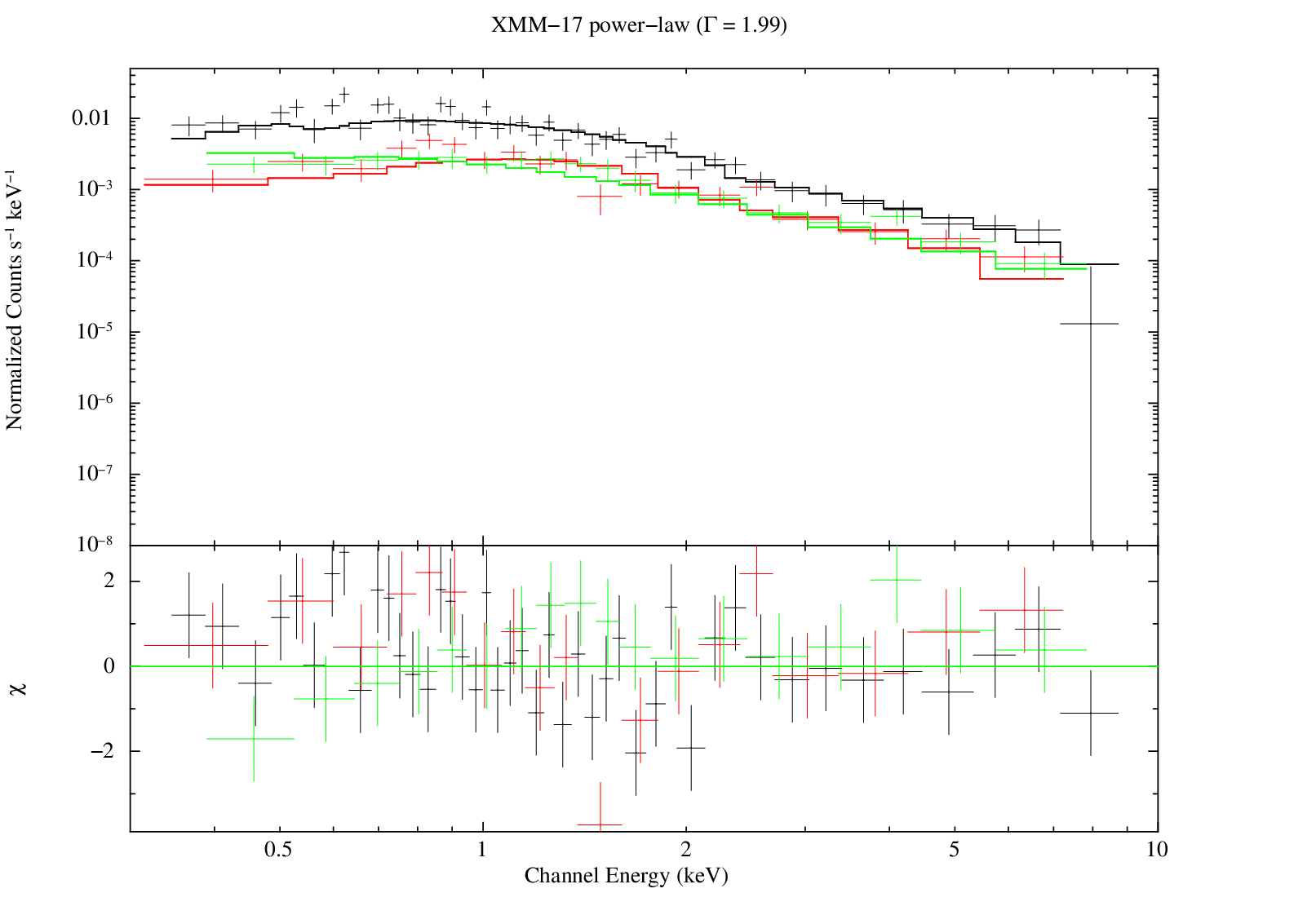}
}
\hspace{0.8cm}
 (h)
{
\label{fig:sub:h}
\includegraphics[width=4.8cm, angle=-90]{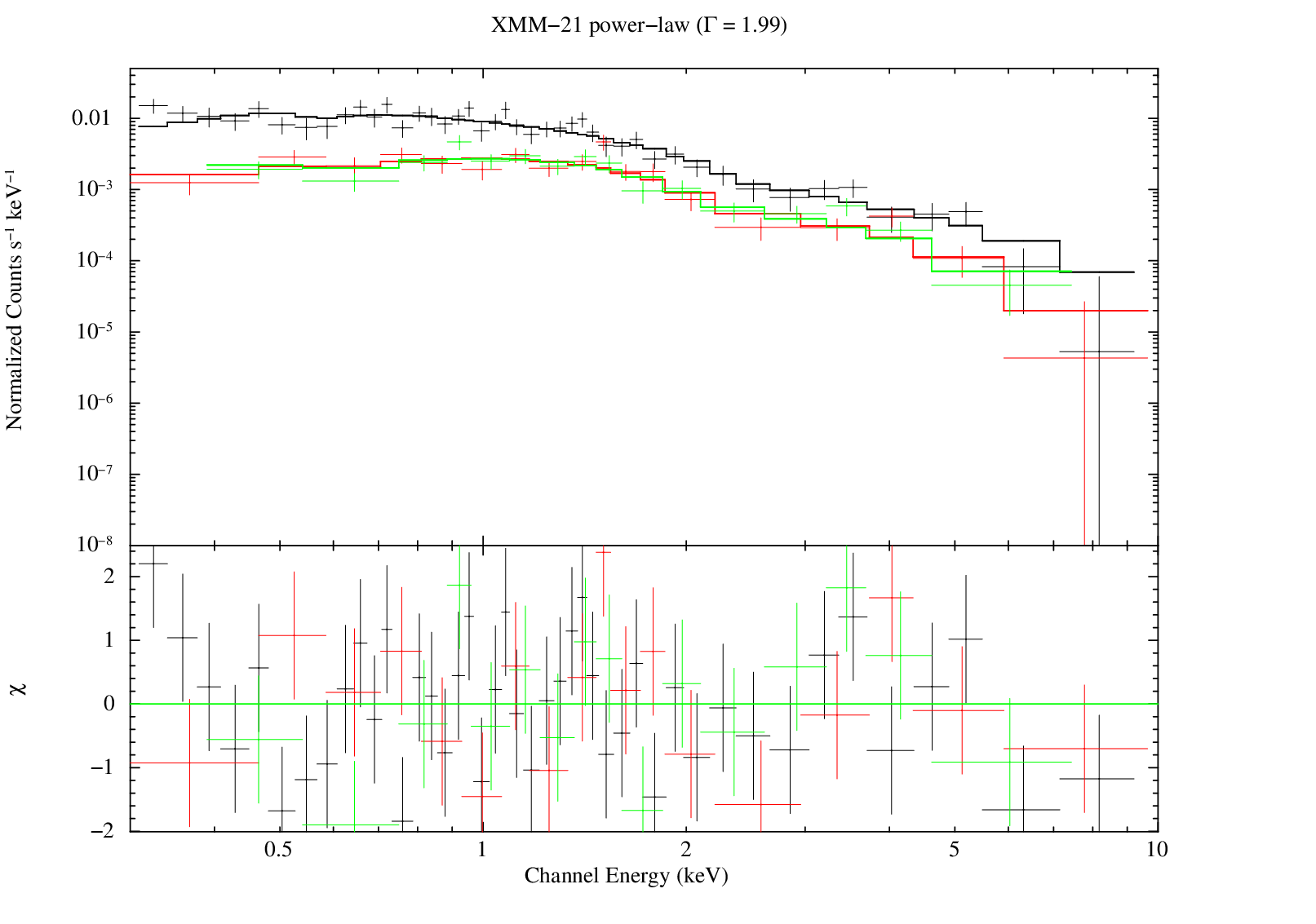}
\caption{The best fitting model, spectra  (upper panels), and the
residuals between the data and the model in standard deviations
(lower panels)  of the sources in NGC 4258. The EPIC-pn data points and the best fitting model are shown in black; data from MOS1 and MOS2  are shown in green and red, respectively.}
}
\label{fig:sub}
\end{figure*}

\clearpage

\setcounter{table}{0}
\begin{deluxetable}{c c c c c l l }
\tablecaption{General properties of the sample galaxies studied in our work.}
\label{table:1}

\tablehead{ \colhead{Galaxy} & \colhead{RA,  Dec.}  & \colhead{Hubble Type$^{1}$} & \colhead{Distance$^{2}$}  & \colhead{Inclination} & \colhead{${n_{H}}^{3}$} & \colhead{${D_{25}corr}^{4}$}  \\
       & (J2000)        &             &  (Mpc)   & (deg.)& $(10^{20}$cm$^{-2}$)   &(arcmin) } 
\startdata
NGC 4395 & 12:25:48.9 +33:32:47.8 & SA(s)m     & 4.2 & 38 & 1.33 & 13.3  \\
NGC 4736 & 12:50:52.6 +41:07:09.3 & (R)SA(r)ab & 4.3 & 33 & 1.43 & 11.8   \\
NGC 4258 & 12:18:57.6 +47:18:13.4 & SAB(s)bc   & 7.7 & 71 & 1.20 & 14.2 \\
\enddata

\tablenotetext{1}{Winter et al. (2006)} 
\tablenotetext{2}{Swartz et al. (2004), Winter et al. (2006)}
\tablenotetext{3}{Column density, obtained from the Web version of the n$_H$ FTOOL.}
\tablenotetext{4}{D$_{25}$ diameter; defined by the 25 mag per square arcsec brightness level.}
\end{deluxetable}



\setcounter{table}{1}
\begin{deluxetable}{c c c c c }
\tablewidth{0pt}
\tablecaption{Locations of luminous sources (as defined in the text) with respect to the galaxy centers and cross identifications, for the sources in three galaxies studied.}             
\label{table:1}

\tablehead{\colhead{Galaxy} & \colhead{Source}  & \colhead{Separation (arcmin.)} & \colhead{ROSAT ID}  & \colhead{CHANDRA ID}  }

\startdata

NGC 4395 & XMM-2 & 2.9& P1$^{a}$ &   \\
         & XMM-5 & 2.1 & P2$^{a}$ &   CXOU J122539.5+333204$^{d}$ \\
         & XMM-6 & 0.7 & P3$^{a}$ &   CXOU J122549.063+333201.82$^{e}$\\
         & XMM-10 & 2.0 & P4$^{a}$ &  CXOU J122547.287+333447.75$^{e}$ \\
         & XMM-23 & 1.9 & &  CXOU J122545.271+333103.40$^{e}$\\

	 &  &  &  &   \\
NGC 4736 & XMM-2 & 0.9 & & CXOU J125048.598+410742.49$^{e}$ \\
         & XMM-12 & 2.3 & X-7$^{b}$ & CXOU J125047.618+410512.26$^{e}$\\
         & XMM-18 & 2.1 & X-13$^{b}$ & CXOU J125104.4+410725$^{f}$\\

	 &  &  &  &   \\
NGC 4258 & XMM-2 & 3.1 & &  \\
         & XMM-3 & 2.1& P16$^{c}$ & CXOU J121857.8+471607$^{d}$ \\
         & XMM-6 & 2.4 &  & CXOU J121843.8+471731$^{d}$ \\  
         & XMM-8 & 1.5& P14$^{c}$ & \\
         & XMM-10 & 6.4 & P13$^{c}$ & \\
         & XMM-16 & 3.3&  & CXOU J121839.317+471919.27$^{e}$\\
         & XMM-17 & 2.0 &  & CXOU J121849.481+471646.42$^{e}$\\
         & XMM-21 & 3.2 & P15$^{c}$ & CXOU J121856.419+472125.56$^{e}$\\

\enddata

\tablenotetext{a}{The 2RXP Catalog 2000. $^{b}$Roberts et al. 1999. $^{c}$Pietsch et al. 1993. $^{d}$Swartz et al. 2004. $^e$Liu 2011. $^f$From present work.}

\end{deluxetable}


\setcounter{table}{2}
\begin{deluxetable}{c c c  l l }
\tablewidth{0pt}
\tablecaption{The log of {\it XMM-Newton} observations, for the three galaxies studied.}
\label{table:1}      

\tablehead{\colhead{Galaxy} &  \colhead{ObsId} & \colhead{Obs. Date} & \colhead{Exposure$^{1}$}  & \colhead{Mode$^{2}$/Filter} \\
       &    &    & (s) & }
\startdata
NGC 4395 & 0112521901 & 2002.05.31 & 13978 & FF/Thin \\
         & \textbf{0142830101} & \textbf{2003.11.30} & \textbf{89000} & \textbf{FF/Medium} \\  
\\
NGC 4736 & 0094360701 & 2002.06.26 & 17046 & FF/Medium \\
         & \textbf{0404980101} & \textbf{2006.11.27} & \textbf{37245} & \textbf{FF/Thin} \\ 
\\
NGC 4258 & 0110920101 & 2000.12.08 & 16548 & FF/Medium \\
         & 0059140101 & 2001.05.06 & 9488 &  FF/Medium \\
         & 0059140201 & 2001.06.17 & 10058 & FF/Medium \\
         & 0059140401 & 2001.12.17 & 11798 & FF/Medium \\
         & \textbf{0059140901} & \textbf{2002.05.22} & \textbf{13646} & \textbf{FF/Medium} \\
         & \textbf{0400560301} & \textbf{2006.11.17} & \textbf{62635} & \textbf{FF/Medium} \\

\enddata     
\tablenotetext{*}{We used only the observations  highlighted in {\bf bold} for the spectral analysis.}
\tablenotetext{1}{A net good time for EPIC-pn after most prominent flares were cut.}
\tablenotetext{2}{FF: Full Frame.}
   
\end{deluxetable}


\setcounter{table}{3}
\begin{deluxetable} {c c c c c c c c}
\tablecaption{Parameters of the sources in NGC 4395 from XMM-Newton observation.}
\tablewidth{0pt}
\tabletypesize{\tiny}

\tablehead{\colhead{Source}  & \colhead{RA(J2000)}  & \colhead{Dec(J2000)}  & \colhead{ML}  & \colhead{Count Rate}  & \colhead{Flux}  & \colhead{Luminosity}   \\
       & (hh:mm:ss.ss) & (dd:mm:ss.s) & &ct s$^{-1}$ & erg cm$^{-2}$s$^{-1}$ & erg s$^{-1}$ }  


\startdata

XMM-17 & 12:25:20.49 & +33:33:00.8 & 1.22E+03 & 1.53E-02$\pm$6.3E-04 & 3.47E-14$\pm$1.7E-15 & 6.24E+37$\pm$3.1E+36   \\

 &  &  &  &  &  &   \\
XMM-55 & 12:25:22.18 & +33:33:16.9 & 1.92E+02 & 5.81E-03$\pm$4.5E-04 & 1.49E-14$\pm$1.2E-15 & 2.68E+37$\pm$2.1E+36   \\

 &  &  &  &  &  &   \\
XMM-184 & 12:25:29.06 & +33:36:11.4 & 2.05E+01 & 1.83E-03$\pm$3.2E-04 & 4.39E-15$\pm$9.4E-16 & 7.90E+36$\pm$1.7E+36   \\

 &  &  &  &  &  &   \\
XMM-107 & 12:25:29.26 & +33:29:01.3 & 8.88E+01 & 3.65E-03$\pm$3.8E-04 & 7.82E-15$\pm$1.0E-15 & 1.40E+37$\pm$1.8E+36   \\

 &  &  &  &  &  &   \\
XMM-43 & 12:25:34.45 & +33:32:11.8 & 2.32E+02 & 4.65E-03$\pm$3.5E-04 & 1.32E-14$\pm$1.0E-15 & 2.37E+37$\pm$1.8E+36  \\

 &  &  &  &  &  &   \\
XMM-113 & 12:25:35.28 & +33:28:31.5 & 7.00E+01 & 3.0E-03$\pm$4.2E-04 & 7.35E-15$\pm$1.0E-15 & 1.32E+37$\pm$1.8E+36   \\

 &  &  &  &  &  &   \\
XMM-147 & 12:25:39.25 & +33:33:49.8 & 4.58E+01 & 1.42E-03$\pm$2.0E-04 & 3.84E-15$\pm$6.2E-16 & 6.91E+36$\pm$1.1E+36   \\

 &  &  &  &  &  &   \\
XMM-5 & 12:25:39.55 & +33:32:04.1 & 5.35E+03 & 3.63E-02$\pm$7.5E-04 & 8.23E-14$\pm$2.0E-15 & 1.48E+38$\pm$3.6E+36   \\

 &  &  &  &  &  &   \\
XMM-99 & 12:25:41.09 & +33:31:09.4 & 3.62E+01 & 1.97E-03$\pm$2.7E-04 & 4.82E-15$\pm$7.3E-16 & 8.67E+36$\pm$1.3E+36 \\

 &  &  &  &  &  &   \\
XMM-109 & 12:25:43.63 & +33:35:07.0 & 6.69E+01 & 1.52E-03$\pm$2.0E-04 & 3.30E-15$\pm$5.5E-16 & 5.94E+36$\pm$9.9E+35   \\

 &  &  &  &  &  &   \\
XMM-145 & 12:25:43.79 & +33:28:54.4 & 6.23E+01 & 2.59E-03$\pm$3.0E-04 & 5.75E-15$\pm$8.2E-16 & 1.03E+37$\pm$1.4E+36  \\

 &  &  &  &  &  &   \\
XMM-16 & 12:25:43.95 & +33:30:00.2 & 1.08E+03 & 1.32E-02$\pm$5.2E-04 & 3.30E-14$\pm$1.5E-15 & 5.94E+37$\pm$2.7E+36   \\ 

 &  &  &  &  &  &   \\
XMM-23 & 12:25:45.11 & +33:31:04.6 & 6.43E+02 & 7.82E-03$\pm$4.0E-04 & 1.95E-14$\pm$1.1E-15 & 3.51E+37$\pm$1.9E+36   \\

 &  &  &  &  &  &   \\
XMM-41 & 12:25:47.05 & +33:36:07.4 & 2.85E+02 & 5.50E-03$\pm$3.6E-04 & 1.38E-14$\pm$1.0E-15 & 2.48E+37$\pm$1.8E+36  \\

 &  &  &  &  &  &   \\
XMM-10 & 12:25:47.20 & +33:34:47.4 & 1.51E+03 & 1.68E-02$\pm$5.5E-04 & 4.00E-14$\pm$1.5E-15 & 7.20E+37$\pm$2.7E+36   \\

 &  &  &  &  &  &   \\
XMM-199 & 12:25:48.70 & +33:28:39.3 & 1.32E+01 & 1.08E-03$\pm$2.1E-04 & 1.74E-15$\pm$5.1E-16 & 3.13E+36$\pm$9.1E+35   \\

 &  &  &  &  &  &   \\
XMM-6 & 12:25:49.00 & +33:32:03.5 & 2.30E+03 & 4.02E-02$\pm$1.1E-03 & 1.32E-13$\pm$3.3E-15 & 2.37E+38$\pm$5.9E+36  \\

 &  &  &  &  &  &   \\
XMM-60 & 12:25:54.08 & +33:29:05.9 & 1.82E+02 & 4.52E-03$\pm$3.4E-04 & 1.03E-14$\pm$9.7E-16 & 1.85E+37$\pm$1.7E+36   \\

 &  &  &  &  &  &   \\
XMM-32 & 12:25:54.44 & +33:30:47.0 & 6.10E+02 & 1.29E-02$\pm$5.8E-04 & 3.20E-14$\pm$1.6E-15 & 5.76E+37$\pm$2.8E+36   \\

 &  &  &  &  &  &   \\
XMM-114 & 12:25:57.32 & +33:29:47.1 & 5.50E+01 & 2.50E-03$\pm$3.1E-04 & 6.17E-15$\pm$9.2E-16 & 1.11E+37$\pm$1.6E+36 \\

 &  &  &  &  &  &   \\
XMM-66 & 12:25:57.35 & +33:30:39.8 & 8.77E+01 & 2.31E-03$\pm$2.5E-04 & 4.42E-15$\pm$7.2E-16 & 7.95E+36$\pm$1.3E+36   \\

 &  &  &  &  &  &   \\
XMM-155 & 12:25:57.98 & +33:36:09.8 & 1.34E+01 & 1.05E-03$\pm$2.2E-04 & 2.88E-15$\pm$6.2E-16 & 5.18E+36$\pm$1.1E+36   \\

 &  &  &  &  &  &   \\
XMM-100 & 12:25:58.87 & +33:28:23.7 & 9.08E+01 & 3.68E-03$\pm$3.6E-04 & 8.71E-15$\pm$1.0E-15 & 1.56E+37$\pm$1.8E+36   \\

 &  &  &  &  &  &   \\
XMM-24 & 12:25:59.79 & +33:33:22.0 & 5.43E+01 & 5.82E-03$\pm$5.6E-04 & 1.35E-14$\pm$1.5E-15 & 2.43E+37$\pm$2.7E+36   \\

 &  &  &  &  &  &   \\
XMM-2 & 12:26:01.42 & +33:31:32.0 & 1.34E+05 & 4.15E-01$\pm$2.3E-03 & 9.12E-13$\pm$6.2E-15 & 1.64E+39$\pm$1.1E+37  \\
         
 &  &  &  &  &  &   \\
XMM-86 & 12:26:04.10 & +33:32:50.6 & 6.42E+01 & 1.85E-03$\pm$3.4E-04 & 7.96E-15$\pm$9.7E-16 & 1.43E+37$\pm$1.7E+36  \\

 &  &  &  &  &  &   \\
XMM-201 & 12:26:13.79 & +33:33:45.9 & 1.36E+01 & 1.47E-03$\pm$2.6E-04 & 3.79E-15$\pm$7.4E-16 & 6.82E+36$\pm$1.3E+36  \\

 &  &  &  &  &  &   \\
XMM-215 & 12:26:15.83 & +33:33:48.0 & 1.44E+01 & 1.28E-03$\pm$2.4E-04 & 2.80E-15$\pm$6.5E-16 & 5.04E+36$\pm$1.1E+36   \\

 &  &  &  &  &  &   \\
XMM-135 & 12:26:16.42 & +33:30:18.2 & 3.74E+01 & 1.81E-03$\pm$2.6E-04 & 4.05E-15$\pm$7.3E-16 & 7.29E+36$\pm$1.3E+36   \\
\enddata 
        
\tablecomments{The sources are ordered by increasing right ascension (RA). Column 1: Source ID; Column 2-3: source coordinates; Column 4: likelihood of existence; Column 5: integrated EPIC pn and MOS  count rates and errors in the  0.2-12 keV band;  Column 6-7: Flux and Luminosity in the 0.2-12 keV band.}
\end{deluxetable}

\setcounter{table}{4}
\begin{deluxetable}{c c c c c c c c c c c c c c c }

\tablecaption{Parameters of the sources in NGC 4736 from XMM-Newton observation.}     

\tablewidth{0pt}

\tabletypesize{\tiny}

\label{table:1}      
  
\tablehead{\colhead{Source}  & \colhead{RA(J2000)}  & \colhead{Dec(J2000)}  & \colhead{ML}  & \colhead{Count Rate}  & \colhead{Flux}  & \colhead{Luminosity}   \\
       & (hh:mm:ss.ss) & (dd:mm:ss.s) & &ct s$^{-1}$ & erg cm$^{-2}$s$^{-1}$ & erg s$^{-1}$ } 
\startdata
XMM-58 & 12:50:27.02 & +41:05:07.5 & 7.49E+01 & 6.47E-03$\pm$9.1E-04 & 3.30E-15$\pm$1.2E-15 & 6.89E+36$\pm$2.5E+36  \\

 &  &  &  &  &  &   \\
XMM-31 & 12:50:33.10 & +41:05:12.8 & 1.48E+02 & 9.58E-03$\pm$8.5E-04 & 2.30E-14$\pm$2.4E-15 & 4.80E+37$\pm$5.0E+36  \\

 &  &  &  &  &  &   \\
XMM-73 & 12:50:34.93 & +41:07:43.7 & 3.16E+01 & 3.94E-03$\pm$5.7E-04 & 7.76E-15$\pm$1.4E-15 & 1.62E+37$\pm$2.9E+36  \\

 &  &  &  &  &  &   \\
XMM-19 & 12:50:35.25 & +41:10:26.4 & 2.42E+02 & 1.086E-02$\pm$8.3E-04 & 2.92E-14$\pm$2.4E-15 & 6.10E+37$\pm$5.0E+36   \\

 &  &  &  &  &  &   \\
XMM-63 & 12:50:35.73 & +41:06:43.9 & 3.54E+01 & 3.81E-03$\pm$5.8E-04 & 1.08E-14$\pm$1.7E-15 & 2.25E+37$\pm$3.5E+36   \\

 &  &  &  &  &  &   \\
XMM-76 & 12:50:37.69 & +41:10:37.9 & 3.48E+01 & 3.80E-03$\pm$5.7E-04 & 9.42E-15$\pm$1.6E-15 & 1.96E+37$\pm$3.3E+36   \\

 &  &  &  &  &  &   \\
XMM-12 & 12:50:47.72 & +41:05:10.8 & 5.28E+02 & 1.95E-02$\pm$1.0E-03 & 4.45E-15$\pm$2.8E-15 & 9.30E+36$\pm$5.8E+36  \\

 &  &  &  &  &  &   \\
XMM-41 & 12:50:48.00 & +41:08:35.1 & 2.25E+01 & 4.29E-03$\pm$6.6E-04 & 9.24E-15$\pm$1.7E-15 & 1.93E+37$\pm$3.5E+36   \\

 &  &  &  &  &  &   \\
XMM-2 & 12:50:48.64 & +41:07:40.5 & 4.77E+04 & 5.62E-01$\pm$4.4E-03 & 1.31E-12$\pm$1.2E-14 & 2.73E+39$\pm$2.5E+37 \\

 &  &  &  &  &  &   \\
XMM-72 & 12:50:50.14 & +41:11:29.4 & 2.39E+01 & 3.00E-03$\pm$5.3E-04 & 8.41E-15$\pm$1.5E-15 & 1.75E+37$\pm$3.1E+36   \\         

 &  &  &  &  &  &   \\
XMM-3 & 12:50:50.27 & +41:07:10.1 & 2.63E+03 & 1.39E-01$\pm$2.8E-03 & 3.31E-13$\pm$7.9E-15 & 6.91E+39$\pm$1.6E+37   \\

 &  &  &  &  &  &   \\
XMM-36 & 12:50:52.57 & +41:02:52.7 & 1.92E+02 & 9.27E-03$\pm$8.5E-04 & 2.18E-14$\pm$2.3E-15 & 4.55E+37$\pm$4.8E+36 \\

 &  &  &  &  &  &   \\
XMM-24 & 12:50:53.00 & +41:08:43.5 & 1.06E+01 & 3.41E-03$\pm$6.8E-04 & 8.63E-15$\pm$1.8E-15 & 1.80E+37$\pm$3.7E+36  \\

 &  &  &  &  &  &   \\
XMM-56 & 12:50:53.48 & +41:04:55.2 & 4.19E+01 & 4.40E-03$\pm$5.8E-04 & 9.79E-15$\pm$1.5E-15 & 2.04E+37$\pm$3.1E+36   \\

 &  &  &  &  &  &   \\
XMM-6 & 12:50:53.67 & +41:06:34.2 & 3.57E+02 & 7.34E-02$\pm$3.5E-03 & 1.67E-13$\pm$9.6E-15 & 3.49E+38$\pm$2.0E+37   \\

 &  &  &  &  &  &   \\
XMM-39 & 12:50:54.24 & +41:09:06.4 & 4.94E+01 & 3.35E-03$\pm$4.6E-04 & 5.70E-15$\pm$1.1E-15 & 1.19E+37$\pm$2.2E+36   \\

 &  &  &  &  &  &   \\
XMM-4 & 12:50:56.10 & +41:06:55.9 & 3.55E+03 & 4.16E-01$\pm$7.4E-03 & 9.44E-13$\pm$2.0E-14 & 1.97E+39$\pm$4.1E+37  \\ 

 &  &  &  &  &  &   \\
XMM-11 & 12:50:59.87 & +41:02:53.8 & 5.45E+02 & 8.49E-03$\pm$6.4E-04 & 3.70E-15$\pm$9.6E-15 & 7.73E+36$\pm$2.0E+37  \\  
  
 &  &  &  &  &  &   \\
XMM-44 & 12:51:02.00 & +41:03:04.7 & 9.74E+01 & 5.98E-03$\pm$6.2E-04 & 1.51E-14$\pm$1.8E-15 & 3.15E+37$\pm$3.7E+36   \\

 &  &  &  &  &  &   \\
XMM-18 & 12:51:04.43 & +41:07:26.0 & 2.10E+02 & 9.90E-03$\pm$7.6E-04 & 2.67E-14$\pm$2.2E-15 & 5.58E+37$\pm$4.5E+36   \\

 &  &  &  &  &  &   \\
XMM-27 & 12:51:09.01 & +41:02:57.8 & 2.37E+02 & 9.96E-03$\pm$7.7E-04 & 2.63E-14$\pm$2.2E-15 & 5.49E+37$\pm$4.5E+36  \\
\enddata 

\tablecomments{Definitions are the same as in Table 4.}

\end{deluxetable}

\setcounter{table}{5}
\begin{deluxetable}{c c c c c c c c c c c c c c }

\tablecaption{Parameters of the sources in NGC 4258 from XMM-Newton observation.}   

\label{table:1}      
\tablewidth{0pt}

\tabletypesize{\tiny}

\tablehead{\colhead{Source}  & \colhead{RA(J2000)}  & \colhead{Dec(J2000)}  & \colhead{ML}  & \colhead{Count Rate}  & \colhead{Flux}  & \colhead{Luminosity}   \\
       & (hh:mm:ss.ss) & (dd:mm:ss.s) & &ct s$^{-1}$ & erg cm$^{-2}$s$^{-1}$ & erg s$^{-1}$ } 

\startdata
XMM-16 & 12:18:39.38 & +47:19:19.1 & 2.37E+03 & 5.02E-02$\pm$1.4E-03 & 1.23E-13$\pm$3.9E-15 & 7.19E+38$\pm$2.2E+37   \\

 &  &  &  &  &  &   \\
XMM-93 & 12:18:42.24 & +47:23:46.9 & 4.19E+01 & 3.15E-03$\pm$5.2E-04 & 7.33E-15$\pm$1.5E-15 & 4.28E+37$\pm$8.7E+36   \\

 &  &  &  &  &  &   \\
XMM-6 & 12:18:43.94 & +47:17:31.7 & 6.91E+03 & 8.78E-02$\pm$1.6E-03 & 2.16E-13$\pm$4.5E-15 & 1.26E+39$\pm$2.6E+37   \\

 &  &  &  &  &  &   \\
XMM-25 & 12:18:45.44 & +47:18:32.1 & 1.28E+03 & 1.22E-01$\pm$3.4E-03 & 3.06E-13$\pm$9.7E-15 & 1.79E+39$\pm$5.6E+37   \\

 &  &  &  &  &  &   \\
XMM-10 & 12:18:45.48 & +47:24:20.5 & 5.06E+03 & 7.96E-02$\pm$1.8E-03 & 2.15E-13$\pm$5.6E-15 & 1.25E+39$\pm$3.2E+37   \\

 &  &  &  &  &  &   \\
XMM-33 & 12:18:45.78 & +47:18:07.5 & 3.44E+01 & 6.05E-03$\pm$7.7E-04 & 1.40E-14$\pm$2.0E-15 & 8.19E+37$\pm$1.1E+37   \\

 &  &  &  &  &  &   \\
XMM-27 & 12:18:46.00 & +47:23:12.9 & 7.49E+02 & 1.94E-02$\pm$9.2E-04 & 4.45E-14$\pm$2.5E-15 & 2.60E+38$\pm$1.4E+37   \\

 &  &  &  &  &  &   \\
XMM-26 & 12:18:47.29 & +47:20:22.0 & 4.09E+02 & 1.47E-02$\pm$8.2E-04 & 3.59E-14$\pm$2.2E-15 & 2.10E+38$\pm$1.2E+37   \\

 &  &  &  &  &  &   \\
XMM-2 & 12:18:47.57 & +47:20:53.8 & 3.26E+04 & 1.64E-01$\pm$4.1E-03 & 4.44E-13$\pm$1.2E-14 & 2.59E+39$\pm$7.0E+37  \\

 &  &  &  &  &  &   \\
XMM-67 & 12:18:48.46 & +47:22:26.1 & 8.41E+01 & 5.10E-03$\pm$5.4E-04 & 1.02E-14$\pm$1.4E-15 & 5.96E+37$\pm$8.1E+36   \\

 &  &  &  &  &  &   \\
XMM-17 & 12:18:49.55 & +47:16:46.6 & 1.41E+03 & 3.29E-02$\pm$1.0E-03 & 8.13E-14$\pm$3.0E-15 & 4.75E+38$\pm$1.7E+37   \\

 &  &  &  &  &  &   \\
XMM-38 & 12:18:53.68 & +47:17:24.5 & 8.64E+01 & 7.18E-03$\pm$7.4E-04 & 1.98E-14$\pm$2.1E-15 & 1.15E+38$\pm$1.2E+37   \\

 &  &  &  &  &  &   \\
XMM-8 & 12:18:54.72 & +47:16:49.4 & 4.93E+03 & 9.35E-02$\pm$1.8E-03 & 2.43E-13$\pm$5.6E-15 & 1.42E+39$\pm$3.2E+37   \\

 &  &  &  &  &  &   \\
XMM-104 & 12:18:54.79 & +47:13:59.7 & 2.24E+01 & 2.23E-03$\pm$4.6E-04 & 6.46E-15$\pm$1.4E-15 & 3.77E+37$\pm$8.1E+36   \\

 &  &  &  &  &  &   \\
XMM-51 & 12:18:55.14 & +47:23:02.5 & 1.90E+02 & 8.50E-03$\pm$6.6E-04 & 2.13E-14$\pm$1.8E-15 & 1.24E+38$\pm$1.0E+37   \\

 &  &  &  &  &  &   \\
XMM-82 & 12:18:56.27 & +47:15:08.1 & 5.03E+01 & 4.15E-03$\pm$5.1E-04 &  8.74E-15$\pm$1.3E-15 & 5.11E+37$\pm$7.6E+36   \\

 &  &  &  &  &  &   \\
XMM-21 & 12:18:56.34 & +47:21:25.6 & 1.46E+03 & 3.00E-02$\pm$1.0E-03 & 7.15E-14$\pm$2.8E-15 & 4.18E+38$\pm$1.6E+37  \\

 &  &  &  &  &  &   \\
XMM-41 & 12:18:57.93 & +47:16:56.7 & 1.70E+01 & 2.68E-03$\pm$4.4E-04 & 5.94E-15$\pm$1.1E-15 & 3.47E+37$\pm$6.4E+36   \\         

 &  &  &  &  &  &   \\
XMM-3 & 12:18:57.94 & +47:16:07.7 & 1.36E+04 & 1.33E-01$\pm$1.9E-03 & 3.37E-13$\pm$5.5E-15 & 1.97E+39$\pm$3.2E+37  \\

 &  &  &  &  &  &   \\
XMM-20 & 12:18:57.96 & +47:19:07.4 & 9.25E+01 & 1.40E-02$\pm$1.2E-03 & 3.13E-14$\pm$3.2E-15 & 1.83E+38$\pm$1.8E+37   \\

 &  &  &  &  &  &   \\
XMM-91 & 12:19:04.35 & +47:14:53.3 & 5.58E+01 & 3.36E-03$\pm$5.1E-04 & 8.06E-15$\pm$1.4E-15 & 4.71E+37$\pm$8.1E+36   \\

 &  &  &  &  &  &   \\
XMM-54 & 12:19:04.90 & +47:16:03.9 & 1.53E+02 & 5.02E-03$\pm$4.8E-04 & 9.43E-15$\pm$1.3E-15 & 5.51E+37$\pm$7.6E+36   \\

 &  &  &  &  &  &   \\
XMM-102 & 12:19:13.23 & +47:14:11.2 & 2.59E+01 & 2.87E-03$\pm$4.2E-04 & 7.90E-15$\pm$1.3E-15 & 4.62E+37$\pm$7.6E+36   \\

 &  &  &  &  &  &   \\
XMM-87 & 12:19:18.59 & +47:16:14.7 & 3.12E+01 & 2.66E-03$\pm$4.0E-04 & 5.57E-15$\pm$1.0E-15 & 3.25E+37$\pm$5.8E+36   \\

\enddata 

\tablecomments{Definitions are the same as in Table 4.}

\end{deluxetable}


\setcounter{table}{6}
\begin{deluxetable}{c c c  l l }  
\tablecaption{The log of {\it Chandra} observations, for the three galaxies studied.}
\label{table:1}      
\tablewidth{0pt} 
                      
\tablehead{\colhead{Galaxy} &  \colhead{ObsId} & \colhead{Obs. Date} & \colhead{Exposure}  & \colhead{Instrument} \\
       &    &    & (ks) & }
\startdata
NGC 4395 & 882 & 2000.06.20 & 20.0 & ACIS-S \\
\\        
NGC 4736 & 808 & 2000.05.13 & 50.0 & ACIS-S \\
         & 9553 & 2008.02.16 & 25.7 & ACIS-I \\ 
\\
NGC 4258 & 350 & 2000.04.17 & 14.5 & ACIS-S \\ 
         & 1618 & 2001.05.28 & 22.0 & ACIS-S\\
         &2340 & 2001.05.29 & 7.0 & ACIS-S\\

\enddata
\end{deluxetable}

\setcounter{table}{7}
\begin{deluxetable}{l l c c c }
\tablecaption{XMM-Newton short-term source variability $\chi^{2}$ test results, for the three galaxies studied.}
\label{table:1}
\tablewidth{0pt}
\tablehead{\colhead{Galaxy} & \colhead{Source}   & \colhead{Bin size} &   \multicolumn{2}{c}{$\chi^{2}$ statistic}     \\
       &          &   (s)    &   $\chi^{2}/dof$       & P$_{\chi^{2}}$ (var)  } 
\startdata
NGC 4395 & XMM-2  &    400    &      254/222   &  -  \\
		 & XMM-5  &    900    &      114/98    &  -    	\\
		 & XMM-6  &    1200   &      100/74    &  97.6 per cent     \\
		 & XMM-10 &    2000   &      29/44     &  -    \\
		 & XMM-23 &    4500   &      23/19     &  - \\
NGC 4736 & XMM-2  &    200    &      880/194   &  $>$99.9 per cent     \\
		 & XMM-12 &    2200   &      17/17     &     -    \\
		 & XMM-18 &    3000   &      9/12      &     -    \\
NGC 4258 & XMM-2  &   900     &      55/27     &   $>$99.9 per cent    \\
		 & XMM-3  &	  700     &     155/91      &   $>$99.9 per cent    \\
		 & XMM-6  &   750     &     223/85      &   $>$99.9 per cent     \\
		 & XMM-8  &   1100    &     94/58       &     99.8 per cent      \\
		 & XMM-10 &   1500    &     47/42       &      -                 \\
		 & XMM-16 &   1500    &     68/42       &       99.3 per cent     \\
		 & XMM-17 &   2500    &     41/25       &       97.7 per cent      \\
		 & XMM-21 &   2000    &     142/32      &       $>$99.9 per cent     \\
\enddata
\end{deluxetable}



\setcounter{table}{8}
\begin{deluxetable}{c c c c c l l l }
\tablecaption{Spectral parameters obtained with one-component model fits for point sources in NGC 4395.}             
\label{table:1}      
\tablewidth{0pt}   
\tablehead{\colhead{Source} & \colhead{model}  & \colhead{N$_{H}$} & \colhead{$\Gamma$} & \colhead{kT}  & \colhead{$\chi^{2}$/dof} & \colhead{F (10$^{-13}$)} & \colhead{L (10$^{38}$)} \\
       &        &$(10^{22})$cm$^{-2}$& & keV & &erg cm$^{-2}$s$^{-1}$&erg s$^{-1}$ }

\startdata

XMM-2 & BBODY & $0.03_{-0.01}^{+0.01}$ & & $0.20_{-0.007}^{+0.006}$ & 253.60/142 & 3.43 & 6.20 \\
         & PL & $0.26_{-0.03}^{+0.03}$  & $4.27_{-0.23}^{+0.26}$ & & 209.13/142 & 6.36 & 11.51  \\
         & DISKBB & $0.07_{-0.01}^{+0.01}$ & & $0.26_{-0.01}^{+0.01}$ & 224.29/142 & 5.23 & 9.46  \\
         & BREMSS & $0.12_{-0.01}^{+0.02}$  & & $0.47_{-0.04}^{+0.04}$ & 206.86/142 & 8.37 & 15.14  \\
  
	 &  &  &  &  &  &  & \\
XMM-5 & BBODY & 0.04 & & 0.26 & 137.22/58 & 0.39 & 0.70 \\
         & PL & 0.31 & 3.62 & & 128.66/58 & 4.27 & 7.72  \\
         & DISKBB & 0.18 & & 0.67 & 129.77/58 & 0.93 & 0.68  \\
         & BREMSS & 0.12 & & 0.35 & 127.49/58 & 0.58 & 1.05  \\
         & $\textbf{VPSHOCK}$ & $\mathbf{0.06_{-0.01}^{+0.02}}$ & & $\mathbf{3.98_{-0.91}^{+2.20}}$ & $\mathbf{106.32/71}$ & $\mathbf{0.52}$ & $\mathbf{1.08}$  \\\

	 &  &  &  &  &  &  & \\
XMM-6 & BBODY & 0.00 & & 0.59 & 188.68/56 & 0.71 & 1.28 \\
         & $\textbf{PL}$ & $\mathbf{0.06_{-0.02}^{+0.02}}$ & $\mathbf{1.42_{-0.09}^{+0.09}}$ & & $\mathbf{63.32/56}$ & $\mathbf{1.45}$ & $\mathbf{2.62}$  \\
         & DISKBB & 0.03 & & $1.89_{-0.35}^{+0.20}$ & 74.58/56 & 1.20 & 2.17  \\
         & BREMSS & $0.04_{-0.01}^{+0.02}$ & & $18.9_{-7.50}^{+18.00}$ & 62.18/56 & 1.37 & 2.47  \\

	 &  &  &  &  &  &  & \\
XMM-10 & BBODY & 0.00 & & 0.47 & 59.16/29 & 0.19 & 0.34 \\
         & $\textbf{PL}$ & $\mathbf{0.15_{-0.05}^{+0.04}}$ & $\mathbf{1.86_{-0.17}^{+0.15}}$ & & $\mathbf{27.12/29}$ & $\mathbf{0.51}$ & $\mathbf{0.92}$  \\
         & DISKBB & 0.02 & & $1.22_{-0.24}^{+0.30}$ & 33.81/29 & 0.31 & 0.56  \\
         & BREMSS & $0.01_{-0.03}^{+0.04}$ & & $5.39_{-1.80}^{+4.03}$ & 31.21/29 & 0.40 & 0.72  \\
	 &  &  &  &  &  &  & \\
XMM-23 & $\textbf{BBODY}$ & $\mathbf{0.28_{-0.01}^{+0.22}}$ & & $\mathbf{0.06_{-0.02}^{+0.03}}$ & $\mathbf{12.94/10}$ & $\mathbf{0.33}$ & $\mathbf{0.60}$ \\
         & PL & $0.17_{-0.14}^{+0.14}$ & $6.87_{-1.83}^{+2.30}$ & & 12.49/10 & 0.98 & 0.17  \\
         & DISKBB & $0.17_{-0.06}^{+0.08}$ &  & $0.09_{-0.02}^{+0.02}$ & 11.57/10 & 0.62 & 0.27 \\
         & BREMSS & $0.21_{-0.05}^{+0.07}$ &  & $0.12_{-0.08}^{+0.10}$ & 11.67/10 & 0.61 & 0.23 \\

\enddata
\tablecomments{The best-fitting model for each source is highlighted in bold.}

\end{deluxetable} 


\setcounter{table}{9}
\begin{deluxetable}{c c c c c c l l } 
\tablecaption{ Spectral parameters obtained using two-component model fits for point sources in NGC 4395} 

\label{table:1}      
\tablewidth{0pt}  

\tablehead{\colhead{Source} & \colhead{model}  & \colhead{N$_{H}$} & \colhead{$\Gamma$} & \colhead{kT}  & \colhead{$\chi^{2}$/dof} & \colhead{F (10$^{-13}$)} & \colhead{L (10$^{38}$)} \\
       &        &$(10^{22})$cm$^{-2}$& & keV & &erg cm$^{-2}$s$^{-1}$&erg s$^{-1}$ }

\startdata

XMM-2    & PL+BBODY & $0.29_{-0.04}^{+0.04}$ & $4.41_{-0.29}^{+0.32}$ & $0.09_{-0.004}^{+2.74}$ & 206.75/140 & 12.8 & 23.16 \\
         & $\textbf{PL+DISKBB}$ & $\mathbf{0.15_{-0.03}^{+0.04}}$ & $\mathbf{3.11_{-0.36}^{+0.43}}$ & $\mathbf{0.19_{-0.02}^{+0.02}}$  & $\mathbf{171.80/140}$ & $\mathbf{ 12.3}$ & $\mathbf{22.26}$ \\
         & PL+MEKAL & $0.18_{-0.02}^{+0.03}$ & $3.80_{-0.15}^{+0.19}$ & $0.64_{-0.07}^{+0.07}$ & 168.06/140 & 23.1 & 41.81 \\

         &           &   &  & & & & \\

XMM-5    & PL+BBODY & $0.12_{-0.05}^{+0.03}$ & $1.71_{-0.56}^{+0.65}$ & $0.20_{-0.03}^{+0.02}$ & 98.02/56 & 0.72 & 1.30 \\
         & PL+DISKBB & 0.28 & 3.44 & 0.06 & 128.78/56 & 0.08 & 0.14 \\
         & PL+MEKAL & $0.42_{-0.08}^{+0.05}$ & $3.40_{-0.20}^{+0.41}$ & $0.17_{-0.02}^{+0.02}$ & 88.59/56 & 5.96 & 10.78 \\ 

	 &           &   &  & & & & \\

XMM-6    & PL+BBODY & $0.05_{-0.04}^{+0.08}$ & $1.40_{-0.09}^{+0.10}$ & 199.36 & 61.73/54 & 1.45 & 2.67 \\
         & PL+DISKBB & $0.04_{-0.03}^{+0.04}$ & $1.34_{-0.42}^{+1.91}$ & $1.25_{-1.25}^{+1.25}$ & 60.30/54 & 1.37 & 2.47 \\  
         & PL+MEKAL & $0.14_{-0.04}^{+0.11}$ & $4.43_{-1.56}^{+1.25}$ & $14.16_{-5.75}^{+20.25}$ & 56.21/54 & 2.36 & 4.27 \\ 

   &           &   &  & & & & \\

XMM-10   & PL+BBODY & $0.16_{-0.06}^{+0.09}$ & $1.95_{-0.50}^{+0.33}$ & $28.88_{-28.90}^{+28.90}$ & 26.80/27 & 0.56 & 1.01 \\ 
         & PL+DISKBB & $0.14_{-0.03}^{+0.06}$ & $1.83_{-0.20}^{+0.30}$ & 0.005 & 26.78/27 & 0.03 & 0.04 \\ 
         & PL+MEKAL & $0.13_{-0.06}^{+0.05}$ & $1.66_{-0.20}^{+0.32}$ & $0.60_{-0.41}^{+0.22}$ & 22.11/27 & 4.94 & 8.94 \\ 

         &           &   &  & & & & \\

XMM-23    & PL+BBODY & $0.30_{-0.11}^{+0.26}$ & $7.17_{-2.05}^{+7.17}$ & $2.19_{-2.19}^{+2.19}$ & 9.94/8 & 18.68 & 33.81 \\ 
          & PL+DISKBB & 1.05 & 0.90 & 0.12 & 19.80/8 & 0.10 & 0.18 \\
          & PL+MEKAL & $0.26_{-0.17}^{+0.26}$ & $6.55_{-7.46}^{+6.55}$ & $0.08_{-0.08}^{+0.18}$ & 10.06/8 & 6.66 & 12.04 \\

\enddata
\tablecomments{The best-fitting model for each source is highlighted in bold.}

\end{deluxetable} 


\setcounter{table}{10}
\begin{deluxetable}{c c c c c l l l l} 
\tablecaption{Relative line intensities and spectroscopic parameters obtained from  TUG observations for the SNR candidate source XMM-5 in NGC 4395}
\label{table:1}
\tablewidth{0pt} 

\tablehead{\colhead{Line} & \colhead{XMM5} }

\startdata
H$\alpha$($\lambda$6563) & 100 \\
SII($\lambda$6716) & 30.8 \\
SII($\lambda$6731) & 11.3 \\
OI($\lambda$6300) & 221.7 \\
OII($\lambda\lambda$7320,7330)& 413 \\
\hline
E$_{(B-V)}$ & 0.02 \\
I(H$\alpha$ )& 2.3E-13 &erg cm$^{-2}$s$^{-1}$ \\
(SII($\lambda$6716)+SII($\lambda$6731))/H$\alpha$&  0.42 \\

\enddata

\end{deluxetable}


\setcounter{table}{11}
\begin{deluxetable}{c c c c c l l l }
\tablecaption{Spectral parameters obtained with one-component model fits for point sources in NGC 4736. }             
\label{table:1}      
    
\tablewidth{0pt}  

\tablehead{\colhead{Source} & \colhead{model}  & \colhead{N$_{H}$} & \colhead{$\Gamma$} & \colhead{kT}  & \colhead{$\chi^{2}$/dof} & \colhead{F (10$^{-13}$)} & \colhead{L (10$^{38}$)} \\
       &        &$(10^{22})$cm$^{-2}$& & keV & &erg cm$^{-2}$s$^{-1}$&erg s$^{-1}$ }

\startdata

XMM-2    & PL & $0.09_{-0.007}^{+0.009}$ & $2.15_{-0.03}^{+0.04}$ & & 768.60/519 & 1.20 & 2.77  \\
         & DISKBB & $<0.008$  & & $0.87_{-0.02}^{+0.02}$ & 695.47/519 & 6.72 & 15.52 \\
         & BREMSS & $0.03_{-0.003}^{+0.004}$ & & $3.07_{-0.16}^{+0.17}$ & 534.87/519 & 8.53 & 19.70 \\
   
    &  &  &  &  &  &  & \\

XMM-12   & PL & $0.04_{-0.04}^{+0.04}$ & $1.47_{-0.23}^{+0.31}$ & & 16.17/19 & 0.54 & 1.24 \\
         & $\textbf{DISKBB}$ & $\mathbf{0.01_{-0.01}^{+0.06}}$ & & $\mathbf{1.44_{-0.19}^{+0.38}}$ & $\mathbf{17.09/19}$ & $\mathbf{0.23}$ & $\mathbf{0.73}$  \\
         & BREMSS & $0.03_{-0.03}^{+0.04}$ & & $12.95_{-7.43}^{+11.43}$ & 15.69/19 & 0.42 & 0.97  \\

   &  &  &  &  &  &  & \\   
    
XMM-18   & $\textbf{PL}$ & $\mathbf{0.09_{-0.09}^{+0.06}}$  & $\mathbf{1.39_{-0.19}^{+0.26}}$ & & $\mathbf{17.42/20}$ & $\mathbf{0.31}$ & $\mathbf{0.71}$\\
         & DISKBB & $0.02_{-0.02}^{+0.05}$ & & $1.97_{-0.98}^{+0.55}$ & 19.80/20 & 0.24 & 0.50  \\
         & BREMSS & $0.07_{-0.07}^{+0.05}$ & & $25.54_{-18.02}^{+25.61}$ & 17.58/20 & 0.29 & 0.66  \\

 &           &   &  & & & & \\
 
\enddata           

\tablecomments{Sources in this galaxy can not be modelled by a BBODY model. The best-fitting model is highlighted in bold. }
\end{deluxetable}

\setcounter{table}{12}
\begin{deluxetable}{c c c c c l l l }
\tablecaption{Spectral parameters obtained with two-component model fits for point sources in NGC 4736.  } 

\label{table:1}      

\tablewidth{0pt}  

\tablehead{\colhead{Source} & \colhead{model}  & \colhead{N$_{H}$} & \colhead{$\Gamma$} & \colhead{kT}  & \colhead{$\chi^{2}$/dof} & \colhead{F (10$^{-13}$)} & \colhead{L (10$^{38}$)} \\
       &        &$(10^{22})$cm$^{-2}$& & keV & &erg cm$^{-2}$s$^{-1}$&erg s$^{-1}$ }

\startdata

XMM-2    & PL+BBODY & $0.03_{-0.006}^{+0.005}$ & $1.92_{-0.06}^{+0.07}$ & $0.44_{-0.03}^{+0.03}$ & 576.06/517 & 8.97 & 20.72 \\
         & $\textbf{PL+DISKBB}$ & $\mathbf{0.02_{-0.006}^{+0.008}}$ & $\mathbf{1.72_{-0.13}^{+0.14}}$ & $\mathbf{0.75_{-0.04}^{+0.06}}$ & $\mathbf{530.81/517}$ & $\mathbf{8.44}$ & $\mathbf{19.49}$ \\
         & PL+MEKAL & $0.06_{-0.009}^{+0.008}$ & $2.05_{-0.06}^{+0.06}$ & $2.55_{-0.35}^{+0.44}$ & 718.94/517 & 9.92 & 24.51 \\

         &           &   &  & & & & \\

XMM-12   & PL+BBODY & $0.05_{-0.05}^{+0.08}$ & $1.81_{-0.86}^{+2.22}$ & $0.98_{-0.98}^{+0.98}$ & 15.48/17 & 0.54 & 1.23 \\
         & PL+DISKBB & $0.09_{-0.08}^{+0.12}$ & $1.57_{-0.32}^{+0.42}$ & $0.04_{-0.04}^{+1.07}$ & 15.52/17 & 0.67 & 1.40 \\
         & PL+MEKAL & $0.05_{-0.05}^{+0.29}$ & $5.64_{-0.22}^{+0.48}$ & 10.70 & 15.77/17 & 1.14 & 2.63 \\

	 &           &   &  & & & & \\

XMM-18   & PL+BBODY & $0.44_{-0.39}^{+0.68}$ & $1.58_{-0.49}^{+0.88}$ & $0.09_{-0.03}^{+0.14}$ & 14.62/18 & 1.24 & 2.59\\
         & PL+DISKBB & $0.54_{-0.44}^{+0.64}$ & $1.66_{-0.54}^{+0.81}$ & $0.10_{-0.01}^{+0.16}$ & 14.78/18 & 3.29 & 6.87 \\
         & PL+MEKAL & $0.31_{-0.12}^{+0.60}$ & $1.45_{-0.34}^{+0.50}$ & $0.21_{-0.06}^{0.24}$ & 13.60/18 & 5.30 & 11.07 \\

	 &           &   &  & & & & \\

\enddata         
\tablecomments{The best-fitting model is highlighted in bold.}

\end{deluxetable}


\setcounter{table}{13}
\begin{deluxetable}{c c c c c l l l } 
\tablecaption{Spectral parameters obtained with one-component model fits for point sources in NGC 4258. }             
\label{table:1}      

\tablewidth{0pt}  

\tablehead{\colhead{Source} & \colhead{model}  & \colhead{N$_{H}$} & \colhead{$\Gamma$} & \colhead{kT}  & \colhead{$\chi^{2}$/dof} & \colhead{F (10$^{-13}$)} & \colhead{L (10$^{38}$)} \\
       &        &$(10^{22})$cm$^{-2}$& & keV & &erg cm$^{-2}$s$^{-1}$&erg s$^{-1}$ }

\startdata
XMM-2$^{*}$ & PL & $0.18_{-0.06}^{+0.07}$ & $1.90_{-0.22}^{+0.21}$ & & 31.80/22 & 3.96 & 23.87 \\
         & $\textbf{DISKBB}$ & $\mathbf{0.02_{-0.02}^{+0.03}}$ & & $\mathbf{1.43_{-0.25}^{+0.30}}$ & $\mathbf{29.28/22}$ & $\mathbf{2.53}$ &$\mathbf{15.25}$ \\
         & BREMSS & $0.11_{-0.04}^{+0.04}$ & & $5.50_{-1.74}^{+2.86}$ & 29.91/22 & 3.14 & 18.96 \\
         
          &           &   &  & & & & \\
XMM-3   & PL & $0.56_{-0.06}^{+0.07}$ & $2.19_{-0.07}^{+0.07}$ & & 297/231 & 6.60 & 39.77 \\
         & $\textbf{DISKBB}$ & $\mathbf{0.27_{-0.03}^{+0.05}}$ & & $\mathbf{1.28_{-0.05}^{+0.05}}$ & $\mathbf{221.14/231}$ & $\mathbf{3.23}$ & $\mathbf{19.46}$ \\
         & BREMSS & $0.41_{-0.04}^{+0.05}$ & & $3.91_{-0.39}^{+0.30}$ & 237.76/231 & 4.23 & 25.49 \\
         &           &   &  & & & & \\
      
XMM-6    & $\textbf{PL}$ & $\mathbf{0.21_{-0.03}^{+0.03}}$ & $\mathbf{1.87_{-0.09}^{+0.09}}$ & & $\mathbf{136.61/142}$ & $\mathbf{2.57}$ & $\mathbf{15.48}$ \\
         & DISKBB & $0.06_{-0.01}^{+0.01}$ & & $1.39_{-0.09}^{+0.10}$ & 161.22/142 & 1.63 & 9.82 \\
         & BREMSS & $0.15_{-0.02}^{+0.02}$ & & $5.60_{-0.84}^{+1.17}$ & 135.84/142 & 2.03 & 12.23 \\
   &           &   &  & & & & \\

XMM-8    & PL & $0.12_{-0.02}^{+0.02}$ & $2.19_{-0.12}^{+0.14}$ & &  125.29/109 & 2.04 & 12.28 \\
         & DISKBB & 0.01 & & 0.83 & 249.70/109 & 1.01 & 6.08 \\
         & BREMSS & $0.05_{-0.01}^{+0.01}$ & & $3.20_{-0.47}^{+0.56}$ & 166.01/109 & 1.37 & 8.25 \\
         &           &   &  & & & & \\

XMM-10    & $\textbf{PL}$ & $\mathbf{0.58_{-0.05}^{+0.05}}$ & $\mathbf{2.80_{-0.14}^{+0.15}}$ & & $\mathbf{87.67/96}$ & $\mathbf{1.54}$ & $\mathbf{9.28}$ \\
          & DISKBB & $0.27_{-0.10}^{+0.10}$ & & $0.84_{-0.09}^{+0.10}$ & 87.16/96 & 1.83 & 11.02 \\
          & BREMSS & $0.39_{-0.03}^{+0.04}$ & & $2.00_{-0.23}^{+0.26}$ & 86.82/96 & 2.61 & 15.73\\
  &           &   &  & & & & \\

XMM-16    & $\textbf{PL}$ & $\mathbf{0.27_{-0.04}^{+0.04}}$ & $\mathbf{2.20_{-0.15}^{+0.18}}$ & & $\mathbf{99.78/76}$ & $\mathbf{1.20}$ & $\mathbf{7.23}$ \\
          & DISKBB & $0.07_{-0.02}^{+0.02}$ & & $0.85_{-0.10}^{+0.10}$ & 146.78/76 & 0.68 & 4.10 \\
          & BREMSS & $0.14_{-0.02}^{+0.03}$ & & $2.59_{-0.45}^{+0.53}$ & 117.96/76 & 0.94 & 5.72 \\
   &           &   &  & & & & \\

XMM-17    & $\textbf{PL}$ & $\mathbf{0.15_{-0.03}^{+0.03}}$ & $\mathbf{1.99_{-0.10}^{+0.12}}$ & & $\mathbf{117.13/80}$ & $\mathbf{0.95}$ & $\mathbf{3.31}$ \\
          & DISKBB & 0.05 & & 1.10 & 169.73/80 & 0.55 & 3.31\\
          & BREMSS & $0.09_{-0.01}^{+0.02}$ & & $4.32_{-0.88}^{+1.09}$ & 131.37/80 & 0.72 & 4.34\\
   &           &   &  & & & & \\

XMM-21    & $\textbf{PL}$ & $\mathbf{0.08_{-0.03}^{+0.03}}$ & $\mathbf{1.99_{-0.09}^{+0.11}}$ & & $\mathbf{85.67/78}$ & $\mathbf{0.34}$ & $\mathbf{2.05}$\\
          & DISKBB & 0.00 & & $0.93_{-0.10}^{+0.10}$ & 100.56/78 & 0.20 & 1.20 \\
          & BREMSS & $0.02_{-0.01}^{+0.02}$ &  & $4.07_{-0.89}^{+1.15}$ & 81.61/78 & 0.29 & 1.74 \\
\enddata  

\tablenotetext{a}{The spectra of this source was obtained by using only the EPIC-pn data}.\\
\tablecomments{Sources in this galaxy can not be modelled by a BBODY model and the best-fitting model for each source is highlighted in bold.}

\end{deluxetable}      


\setcounter{table}{14}
\begin{deluxetable}{c c c c c l l l } 
\tablecaption{ Spectral parameters obtained with two-component model fits for point sources in NGC 4258 } 

\label{table:1}      

\tablewidth{0pt}  

\tablehead{\colhead{Source} & \colhead{model}  & \colhead{N$_{H}$} & \colhead{$\Gamma$} & \colhead{kT}  & \colhead{$\chi^{2}$/dof} & \colhead{F (10$^{-13}$)} & \colhead{L (10$^{38}$)} \\
       &        &$(10^{22})$cm$^{-2}$& & keV & &erg cm$^{-2}$s$^{-1}$&erg s$^{-1}$ }

\startdata
XMM-2$^{*}$  & PL+BBODY & $0.16_{-0.15}^{+0.14}$ & $2.24_{-0.72}^{+0.27}$ & $0.88_{-0.40}^{+0.55}$ & 27.76/20 & 4.05 & 24.42 \\
         & PL+DISKBB & $0.11_{-0.05}^{+0.34}$ & $2.12_{-1.59}^{+1.59}$ & $1.40_{-0.47}^{+1.12}$ & 27.35/20 & 2.80 & 16.88 \\
         & PL+MEKAL & $0.10_{-0.04}^{+0.05}$ & $9.45_{-9.89}^{+9.89}$ & $4.25_{-0.89}^{+1.44}$ & 37.11/20 & 7.89 & 47.57 \\

         &           &   &  & & & & \\

XMM-3    & PL+BBODY & $0.37_{-0.09}^{+0.10}$ & $2.16_{-0.22}^{+0.10}$ & $0.76_{-0.07}^{+0.08}$ & 244.56/229 & 4.39 & 19.76 \\
         & PL+DISKBB & $0.28_{-0.02}^{+0.02}$ & $0.67_{-9.67}^{+19.67}$ & $1.23_{-0.06}^{+0.06}$ & 219.18/229 & 3.28 & 19.76 \\
         & PL+MEKAL & $0.75_{-0.08}^{+0.17}$ & $2.37_{-0.07}^{+0.13}$ & $0.14_{-0.02}^{+0.08}$ & 278.16/229 & 4.82 & 29.05 \\ 

	 &           &   &  & & & & \\

XMM-6    & PL+BBODY & $0.21_{-0.03}^{+0.05}$ & $1.97_{-0.16}^{+0.24}$ & $0.94_{-0.37}^{+0.64}$ & 133.26/140 & 2.30 & 13.85 \\
         & PL+DISKBB & $0.23_{-0.08}^{+0.26}$ & $2.27_{-0.68}^{+2.26}$ & $2.04_{-1.38}^{+2.66}$ & 133.54/140 & 2.42 & 14.58 \\  
         & PL+MEKAL & $0.19_{-0.02}^{+0.19}$ & $1.80_{-0.12}^{+0.21}$ & 4.12 & 135.64/140 & 2.22 & 13.37 \\ 

   &           &   &  & & & & \\

XMM-8    & PL+BBODY & $0.28_{-0.04}^{+0.09}$ & $3.57_{-0.46}^{+0.30}$ & $1.44_{-0.26}^{+0.35}$ & 94.53/107 & 7.84 & 47.24 \\ 
         & PL+DISKBB & $0.12_{-0.02}^{+0.02}$ & $2.19_{-0.12}^{+0.14}$ & 0.00 & 124.87/107 & 2.86 & 17.23 \\ 
         & $\textbf{PL+MEKAL}$ & $\mathbf{0.09_{-0.01}^{+0.02}}$ & $\mathbf{1.95_{-0.13}^{+0.07}}$ & $\mathbf{0.57_{-0.04}^{+0.05}}$ & $\mathbf{105.93/107}$ & $\mathbf{1.96}$ & $\mathbf{11.81}$ \\ 

         &           &   &  & & & & \\

XMM-10    & PL+BBODY & $0.58_{-0.05}^{+0.06}$ & $2.79_{-0.14}^{+0.17}$ & 199.36 & 81.24/92 & 5.19 & 31.27 \\ 
          & PL+DISKBB & $0.56_{-0.24}^{+0.23}$ & $3.05_{-1.02}^{+2.00}$ & $0.99_{-0.41}^{+0.46}$ & 77.65/92 & 5.13 & 30.91 \\
          & PL+MEKAL & $0.58_{-0.08}^{+0.08}$ & $2.94_{-0.29}^{+0.25}$ & $2.95_{-0.51}^{+0.42}$ & 79.81/92 & 1.85 & 11.14 \\
 			
  &           &   &  & & & & \\
XMM-16    & PL+BBODY & $0.29_{-0.05}^{+0.05}$ & $2.61_{-0.35}^{+0.24}$ & 199.36 & 99.31/74 & 2.16 & 13.02 \\
          & PL+DISKBB & $0.27_{-0.08}^{+0.32}$ & $2.47_{-0.08}^{+0.08}$ & $0.98_{-0.05}^{+0.03}$ & 101.47/74 & 1.6 & 9.64 \\
          & PL+MEKAL & $0.17_{-0.04}^{+0.04}$ & $2.09_{-0.14}^{+0.10}$ & $0.65_{-0.22}^{+0.19}$ & 76.69/74 & 0.43 & 2.59 \\

  &           &   &  & & & & \\

XMM-17    & PL+BB & $0.20_{-0.05}^{+0.17}$ & $1.69_{-0.18}^{+0.33}$ & $0.14_{-0.04}^{+0.03}$ & 101.54/78 & 1.12 & 6.75 \\
          & PL+DISKBB & $0.30_{-0.08}^{+0.14}$ & $1.74_{-0.24}^{+0.19}$ & $0.16_{-0.04}^{+0.05}$ & 101.55/78 & 1.83 & 11.03 \\
          & PL+MEKAL & $0.07_{-0.03}^{+0.04}$ & $1.68_{-0.14}^{+0.17}$ & $3.32_{-0.09}^{+0.12}$ & 76.36/78 & 0.42 & 2.53 \\

  &           &   &  & & & & \\

XMM-21    & PL+BB & $0.10_{-0.08}^{+0.05}$ & $2.21_{-0.16}^{+1.03}$ & $0.97_{-0.09}^{+0.12}$ & 84.56/76 & 0.34 & 2.05 \\ 
          & PL+DISKBB & $0.13_{-0.04}^{+0.05}$ & $2.12_{-0.19}^{+0.21}$ & $0.02_{-0.01}^{+0.01}$ & 80.06/76 & 0.32 & 1.92 \\
          & PL+MEKAL & $0.06_{-0.04}^{+0.05}$ & $1.91_{-0.26}^{+0.24}$ & $2.90_{-2.91}^{+2.91}$ & 94.63/76 & 0.33 & 1.98 \\

\enddata        

\tablenotetext{*}{The spectra of this source was obtained by using only the EPIC-pn data.}

\end{deluxetable}      


\end{document}